\documentclass[preprint,aps]{revtex4}
\usepackage{graphics,graphicx,amsmath,amssymb}
\usepackage{chngcntr}
\usepackage[usenames, dvipsnames]{color}
\usepackage[ pdftex, plainpages = false, pdfpagelabels, 
                 pdfpagelayout = useoutlines,
                 bookmarks,
                 bookmarksopen = true,
                 bookmarksnumbered = true,
                 breaklinks = true,
                 linktocpage=all,
                 pagebackref=false,
                 colorlinks = true,
                 linkcolor = BrickRed,
                 urlcolor  = blue,
                 citecolor = BrickRed,
                 anchorcolor = green,
                 hyperindex = true,
                 hyperfigures
                 ]{hyperref} 

\begin{document}
\title{\href{http://necsi.edu/research/social/scienceofpeace.html}{Good Fences:} \\ 
\href{http://necsi.edu/research/social/scienceofpeace.html}{The Importance of Setting Boundaries for Peaceful Coexistence}}
\author{Alex Rutherford} 
\affiliation{\href{http://www.necsi.edu}{New England Complex Systems Institute}\\
238 Main St., Suite 319, Cambridge, MA 02142}
\author{Dion Harmon} 
\affiliation{\href{http://www.necsi.edu}{New England Complex Systems Institute}\\
238 Main St., Suite 319, Cambridge, MA 02142}
\author{Justin Werfel\footnote{current address Wyss Institute, 60 Oxford St, Cambridge, MA 02138}  }
\affiliation{\href{http://www.necsi.edu}{New England Complex Systems Institute}\\
238 Main St., Suite 319, Cambridge, MA 02142}
\author{\\ Shlomiya Bar-Yam} 
\affiliation{\href{http://www.necsi.edu}{New England Complex Systems Institute}\\
238 Main St., Suite 319, Cambridge, MA 02142}
\author{Alexander Gard-Murray}
\affiliation{\href{http://www.necsi.edu}{New England Complex Systems Institute}\\
238 Main St., Suite 319, Cambridge, MA 02142}
\author{Andreas Gros}
\affiliation{\href{http://www.necsi.edu}{New England Complex Systems Institute}\\
238 Main St., Suite 319, Cambridge, MA 02142}
\author{\href{http://necsi.edu/faculty/bar-yam.html}{Yaneer Bar-Yam}}
\affiliation{\href{http://www.necsi.edu}{New England Complex Systems Institute}\\
238 Main St., Suite 319, Cambridge, MA 02142}

\date{May 15, 2011; released October 6, 2011}

\begin{abstract}
We consider the conditions of peace and violence among ethnic groups,
testing a theory designed to predict the locations of violence and 
interventions that can promote peace. Characterizing the model's success in
predicting peace requires examples where peace prevails despite diversity.
Switzerland is recognized as a country of peace, stability and prosperity.
This is surprising because of its linguistic and religious diversity 
that in other parts of the world lead to conflict and violence. 
Here we analyze how peaceful stability is maintained. 
Our analysis shows that peace does not depend on integrated coexistence, 
but rather on well defined topographical and political boundaries 
separating groups. 
Mountains and lakes are an important part of the boundaries between
sharply defined linguistic areas. Political canton and circle (sub-canton) boundaries
often separate religious groups. Where such boundaries do 
not appear to be sufficient, we find that specific aspects of the population distribution 
either guarantee sufficient separation or sufficient mixing to inhibit
intergroup violence according to the quantitative theory of conflict. 
In exactly one region, a porous mountain range
does not adequately separate linguistic groups and violent 
conflict has led to the recent creation of the canton of Jura. Our
analysis supports the hypothesis that violence between groups can be
inhibited by physical and political boundaries. A similar analysis of
the area of the former Yugoslavia shows that during widespread ethnic
violence existing political boundaries did not coincide with the
boundaries of distinct groups, but peace prevailed in specific areas
where they did coincide. The success of peace in Switzerland may
serve as a model to resolve conflict in other ethnically diverse countries 
and regions of the world. 
\end{abstract}

\maketitle

Achieving peace requires a vision of what it looks like. How we imagine peace affects the steps we take and our ability to implement it in diverse locations around the world. Does peace in one place look the same as in another? Is knowledge of the specifics of each conflict necessary to negotiate peace between ethnic groups in conflict? Even if specifics are important, there are broad frameworks that guide our thinking. Recently, we introduced a complex systems theory of ethnic conflict that describes the conflicts in areas of the former Yugoslavia and India with high accuracy [1]. In this theory, specific details of history, social and economic conditions are not the primary conditions for peace or conflict. Instead the geographic arrangement of populations is key. Significantly, it points to two distinct conditions that are conducive to peace---well mixed and well separated. The first corresponds to the most commonly striven for framework of an integrated society [2].  The second corresponds to spatial separation, partition and self determination---a historically used but often reviled approach [3].  
Here we consider a more subtle third approach, that of within-state boundaries in which cooperation and separation are both necessary. The success of this approach is of particular importance as the world becomes more connected. As illustrated by the European Union, the role of borders as boundaries is changing. 

\begin{figure}
\includegraphics[width=11cm]{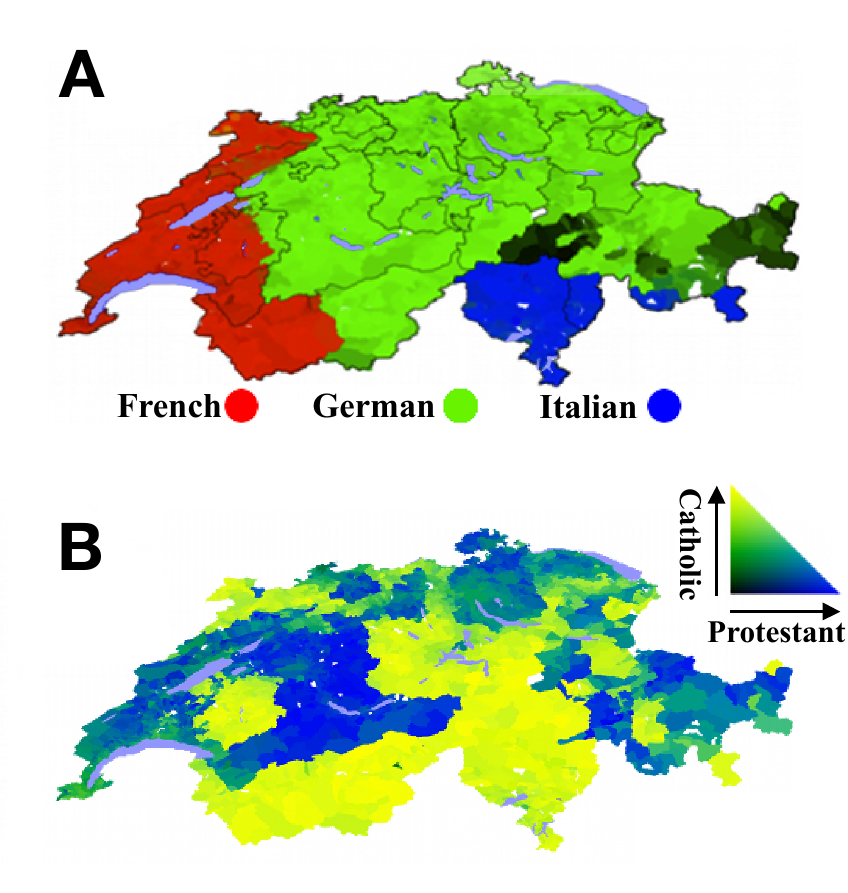}
\caption{\label{fig0} Maps of Switzerland showing the 2000 census 
 proportion of {\bf (A)} linguistic groups, {\bf (B)} Catholic and Protestant (Mercator projection).}
\end{figure}

In order to evaluate the role of within-state boundaries in peace, 
we considered the coexistence of groups in Switzerland. 
Switzerland is known as a country of great stability, without 
major internal conflict despite multiple languages and religions [4,5]. 
Switzerland is not a well-mixed society, it is heterogeneous geographically 
in both language and religion (Fig. \ref{fig0}). 
The alpine topography and the federal system of strong cantons have 
been noted as being relevant to coexistence; their
importance can be seen in Napoleon's statement, after the failure of 
his centralized Helvetic Republic, that ``nature'' had made Switzerland 
a federation [6--8]. But the existence of both alpine and non-alpine
boundaries between groups and the presence of multiple languages and religions within
individual cantons suggest partition is not essential for peaceful
coexistence in Switzerland. In identifying the causes of peace, the
literature has focused on socio-economic and political conditions including
a long tradition of mediation and accommodation, social cleavages that ``cross-cut'' the
population rather than coincide with each other, unwritten and written
rights of proportionality (fairness) and cultural protectionism, a federal 
system with strong sub-national units, a
civil society that fosters unity, direct democracy through frequent
referenda, small size, historical time difference between cleavage in
language and religion, neutrality in international warfare, and
economic prosperity [4--6,9--13]. Geography plays an unclear, presumably
supporting, role in these frameworks. The analysis of coexistence in
Switzerland is also part of a broader debate about whether social and
geographical aspects of federalism promote peace or conflict [15].

In this paper we analyze the geographical distribution of groups in Switzerland
based solely upon the hypothesis that spatial patterns formed by ethnic groups
are predictive of unrest and violence among them [1].  The theory
asserts that highly mixed regions or well-segregated groups are
peaceful, while groups of a certain intermediate geographical size are
likely to engage in violence. While effective separation may be
achieved when group areas are large enough, the model also allows that
topographic or political boundaries may serve as separations to
promote peace [1,16].  Using the case of Switzerland, we test the ability of the theory to
predict peaceful coexistence in the context of internal country
boundaries. Where explicit boundaries do not exist, such as in mixed
cantons where alpine boundaries are absent, violence might be
expected, and the results of the model in these areas serve as a
particularly stringent test of the theory. In most cases violence is
not predicted, consistent with what is found. In one area a significant level of violence
is predicted, and violence is actually observed. The analysis sheds light on
the example of Switzerland as a model for peaceful coexistence.
The former Yugoslavia serves as a contrasting example of 
widespread violence. The theory correctly identifies areas of 
conflict and areas of peace also in the former Yugoslavia.
The precision of the results provides some assurance of the 
usefulness of the theory in planning interventions that might
promote peace in many areas of the world. 

The geographical distribution theory [1] considers type separation 
into geographical domains independent of the specification of the 
individual types---a universality of type behavior in collective violence. 
Violence arises due to the structure of boundaries between
groups rather than as a result of inherent conflicts between the
groups themselves. In this approach, diverse social and economic
causal factors trigger violence when the spatial population structure
creates a propensity to conflict, so that spatial heterogeneity itself
is predictive of local violence. The local ethnic patch size serves as
an ``order parameter,'' a measure of the degree of order of collective
action, to which other aspects of behavior are coupled. The
importance of collective behavior implies that ethnic violence can be
studied in the universal context of collective dynamics, where models
can identify how individual and collective behavior are related.

The analysis is applicable to communal violence and not to criminal
activity or interstate warfare. In highly mixed regions, groups of the
same type are not large enough to develop strong collective
identities, or to identify public spaces as associated with one or
another cultural group.  They are neither imposed upon nor impose upon
other groups, and are not perceived as a threat to the cultural values
or social/political self-determination of other groups. Partial
separation with poorly defined boundaries fosters conflict. Violence
arises when groups are of a geographical size that they are able to
impose cultural norms on public spaces, but where there are still
intermittent violations of these rules due to the overlap of cultural
domains. When groups are larger than the critical size, they typically
form self-sufficient entities that enjoy local sovereignty. Hence, we
expect violence to arise when groups of a certain characteristic size
are formed, and not when groups are much smaller or larger than this
size. The model of violence depends on the distribution of the
population and not on the specific mechanism by which the population
achieves this structure, which may include internally or externally
directed migrations. By focusing on the geographic distribution of the
population, the model seeks a predictor of conflict that can easily be
determined by census. This may work well because geography is an
important aspect of the dimensions of social space, and other aspects
of social behavior (e.g., isolationism, conformity, as well as
violence) are correlated to it.

Physical boundaries such as mountain ranges and lakes or national and
subnational political boundaries that establish local autonomy may
prevent the violations of cultural norms and enable
self-determination, inhibiting the triggers of violence. By creating
autonomous domains of activity and authority, the boundaries shield
groups of the characteristic size from each other when they correspond
with their geographical domains.

Mathematically, evaluation of the model begins by mapping census data
onto a spatial grid. We included the fraction of every population type
on each site. The expected violence is determined by detecting patches
consisting of islands or peninsulas of one type surrounded by
populations of other types. These features are detected by pattern recognition 
using the correlation of the population for each population type with a template that has a
positive center and a negative surround. The template used is based on
a wavelet filter [1,17,18]. Wavelets are designed to obtain a local
measure of the degree to which a certain scale of variation
(wavelength) is present. Outcomes are highly robust, and other
templates give similar results. The diameter of the positive region of
the wavelet, i.e., the size of the local population patches that are
likely to experience violence, is the only essential parameter of the
model. The parameter is to be determined by agreement of the model
with reports of violence, and results were robust to varying the
parameter across a wide range of values. To model the effect of
boundaries, we assume that separate autonomous regions can be analyzed
by including only the populations within each of the autonomous areas
to determine the expected violence. Where boundaries are incomplete,
as might be the case for mountains, lakes and convoluted political
boundaries, we include only the populations that are in line of sight
through gaps or past ends of boundaries to determine the expected
violence within a region. An effective map of populations at each site
is constructed, determined by the orientation of any boundaries
relative to that site. Populations past boundaries of the line of
sight are replaced by neutral populations. The result of the
correlation of population with the wavelet filter is a single value at
every location, the theoretical ``propensity to violence,'' and the
locations of expected violence are obtained by applying a threshold to
that value. The location of groups of a certain size is indicative of
a violence-prone group, but the precise location of violence is not
determined. The proximity of these violence-prone groups to actual
violence is tested by constructing proximity maps. The proximity to
reported violence is correlated to the proximity to violence prone
groups.  The model was validated without boundaries [1] by applying
it to the former Yugoslavia, yielding correlations of up to 0.89. The
results were robust to varying the characteristic length between 18--60
km. Our revised method with fractional population values on every site
gave similar results with correlations of up to 0.87. (Methods are
further described in the Appendix.)

We now consider the linguistic (Fig. \ref{fig1}) and religious
(Fig. \ref{fig2}) groups in Switzerland, each in turn.  Initial
analyses and the sequence of historical boundary formation suggested
considering topographical barriers when discussing language groups,
and political barriers when considering religious groups. The
geography of languages primarily reflects the extent of invasions
prior to the existence of current political boundaries and has
remained stable in most areas for over a thousand years [5]. The
modern state was established afterwards, and religious conflict played
a role in establishing the internal political boundaries [5--7]. Census
data were obtained for 2634 municipalities (communes) in Switzerland
(bfs.admin.ch), yielding a high spatial resolution.

\begin{figure}
\includegraphics[width=16cm]{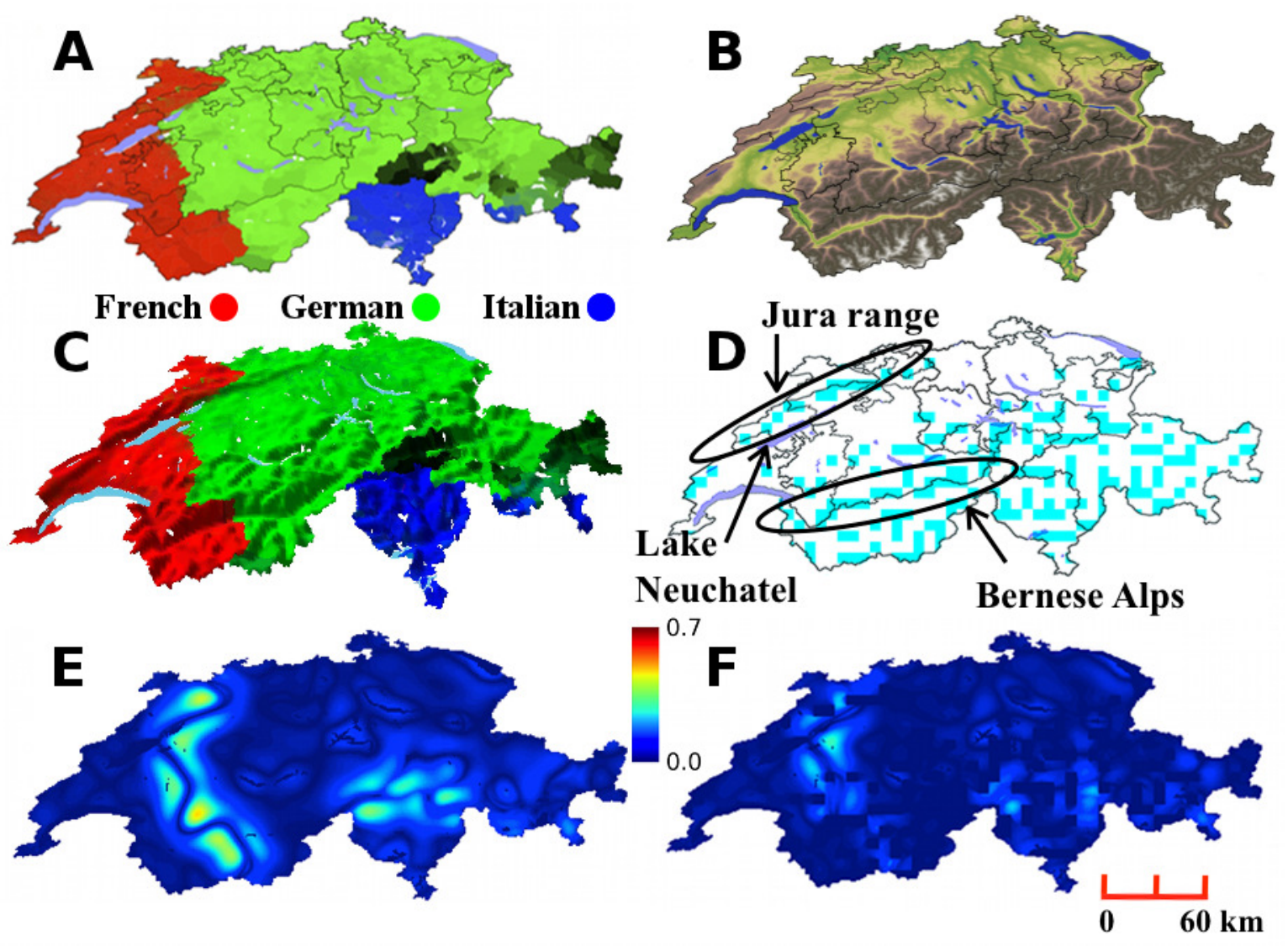}
\caption{\label{fig1} Maps of Switzerland showing {\bf (A)} proportion
  of linguistic groups according to the 2000 census, {\bf (B)}
  elevation within Switzerland, {\bf (C)} overlay of linguistic groups
  onto a digital elevation model, and {\bf (D)} topographical features
  including lakes (blue) and ridges extracted using edge detection
  (cyan). Comparison of calculated propensity (color bar) to violence
  between linguistic groups without {\bf (E)} and with {\bf (F)} the
  inclusion of topographical features as boundaries using a
  characteristic length scale of 24\,km. Mercator projection, except C
  which is the Europe Albers projection. The distance scale is
  approximate.}
\end{figure}

\bigskip
\noindent\textit{Language and topographical barriers} - We study the
three main language groups---German, French and Italian (Fig.
\ref{fig2}A)---which together comprise 91\% of the total population in
the 2000 census (Romansh, the fourth official language, accounts for
less than 2\%). We considered only the effect of physical boundaries
due to lakes and mountain ranges (Fig. \ref{fig2}, B and C). We
determined the presence of topographical boundaries using an edge
detection algorithm on topographical heights (Fig. \ref{fig2}D). This
process identifies where there is a sharp change in height, i.e., a
cliff, or steep incline, that runs for a significant distance forming
a natural boundary.  Elevation data with a spatial resolution of
approximately 91m [19] was coarsened to pixels of size $9.1\times
9.1$\,km. Edges were identified where there was an increase of more
than 1.8\,km in height over a distance of 9.1\,km (11.5$^\circ$) using
a discretized Laplacian differential operator [20] with a mask size of
a single pixel. The conclusions are robust to variations in the
elevation angle (Appendix). Calculations of the propensity to violence
are reported here (Fig. \ref{fig2}, E and F) for the characteristic
length of 24\,km and in the Appendix for a range of characteristic
lengths. Without boundaries, the correlation of the wavelet filter
yields a maximum propensity to violence value of 0.48. With
topographical boundaries the maximum propensity is reduced to
0.30. Between the German and French-speaking areas to the northwest,
the Jura mountain range and Lake Neuchatel, and to the south, the
Bernese Alps, are mitigating boundaries. The interface between Lake
Neuchatel and the Bernese Alps through the canton of Fribourg has no
mitigating boundary, but is almost straight---neither side is
surrounded by the other, so the propensity is low. Between the Italian
and German-speaking areas, the Lepontine Alps dramatically reduce the
calculated propensity.

The Jura range is, however, a porous boundary, and the highest
residual propensity is adjacent to it in the northwest of the canton
of Bern, which, unique in Switzerland, is historically known to be an
area of ``intense'' linguistically-based conflict, including arson,
bombings and other terrorist tactics [13,21]. We obtained a
correlation higher than 0.95 between predicted and reported violence
(Appendix), consistent with the hypothesis of the model. Manifesting
Swiss willingness to create political boundaries, the conflict led to
a referendum, and in 1978 the modern-day canton of Jura was created
out of part of the north of what was then the canton of Bern
[7]. While the conflict underlying the unrest was linguistic, local
votes led to separation by majority religion. However, conflict did
not end, and a proposal to shift the French-speaking Protestant areas
of Bern to join French-speaking Catholic Jura is currently being
considered [22].  Our results suggest that a calculated propensity to
violence of 0.3 should be considered just at the threshold for actual
violence, even under the social and political conditions prevailing in
Switzerland. Remarkably, at this threshold high correlations (above
0.8) also are found in the former Yugoslavia. Thus, similar propensities
for violence in different social contexts result in violence. 

\bigskip

\noindent \textit{Religious Groups and Political Barriers} - The two
main religious groups of Switzerland are Protestant and Catholic. The
Swiss federal political system separates the country into 26
``cantons'' and ``half-cantons'' considered as semi-autonomous
political units (Fig. \ref{fig2}). Moreover, this schema is repeated
within the largest canton by area, Graub\"unden, whose sub-cantonal
divisions called circles (\textit{kreise}) have a distinctive
political autonomy [4,12]. We obtained canton boundaries from mapping
resources (\url{www.gadm.org}, \url{www.toposhop.admin.ch}). Circles
boundaries were identified by district lists (\url{www.gis.gr.ch}). In
the 2000 census, Roman Catholic and Protestant affiliations account
for 77\% of the total population. Less than 8\% subscribe to other
religions, and the remainder have no religious affiliation or did not
specify one. Without boundaries, the maximum calculated propensity to
violence is very high (0.57), and with political borders it is only
0.20. Without Graub\"unden circles, the propensity increases to a
quite high 0.42, still well above the threshold. Because of a 10\%
decline in religious affiliation in recent years, we considered also
the 1990 census, with similar conclusions (Appendix).

\begin{figure}
\includegraphics[width=16cm]{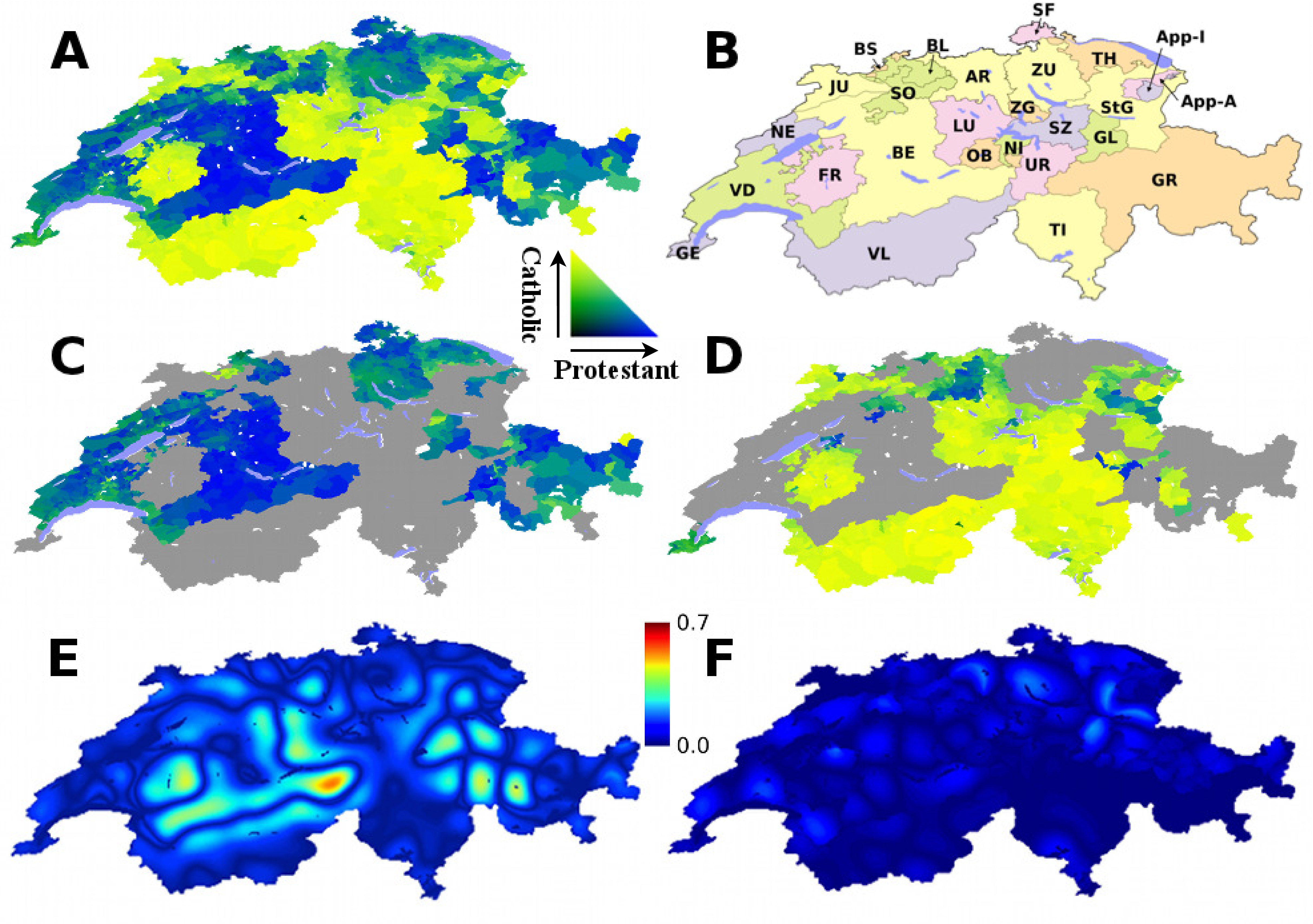}
\caption{\label{fig2} Maps of Switzerland (Mercator projection)
  showing {\bf (A)} proportion of Catholic (yellow) and Protestant
  (blue) according to the 2000 census, {\bf (B)} cantons, {\bf (C)}
  and {\bf (D)} cantons (and Graub\"unden circles) that are majority
  Protestant and Catholic respectively, using the same color map as
  A. Comparison of propensity to violence between religious groups
  without {\bf (E)} and with {\bf (F)} the inclusion of administrative
  boundaries using a characteristic length scale of 24\,km.  Propensity
  value scale is shown by color bar. Canton abbreviations are GE:
  Gen\`eve, SO: Solothurn, ZG: Zug,VL: Valais, BS: Basel-Stadt, GL:
  Glarus, VD: Vaud, BL: Basel-Landschaft, TI: Ticino, NE: Neuchatel,
  AR: Aargau, GR: Graub\"unden, FR: Fribourg, LU: Lucerne, App-A:
  Appenzell-Ausserhoden, BE: Bern, OB: Obwalden, App-I:
  Appenzell-Innerrhoden, JU: Jura, NI: Nidwalden, StG: St. Gallen, UR:
  Uri, SF: Schaffhausen, TH: Thurgau, SZ: Schwyz, ZU: Zurich.}
\end{figure}

The separation of religions by canton is apparent geographically and
historically. In some cases the area of a canton includes small
enclaves embedded in another canton whose majority religion
corresponds to the canton to which they belong. Still, there are
exceptions to the separation of religions by canton. In each case the
geography is sufficient to limit the propensity to violence.  For
example, there is an area of Protestant majority in the far north of
the Catholic canton of Fribourg. It is, however on a long appendage
and therefore is not surrounded by Catholic areas, and so has a low
propensity to violence according to the analysis. Historical evidence
is found in conflict in the 1500s [7]. The Reformation led to cantons
adopting a Protestant or retaining a Catholic identity. A brief war
resulted in a peace treaty that established religious freedom by
canton. The canton Appenzell was split by religious differences into
two ``half-cantons'' Innerrhoden and Ausserrhoden. The political
independence of circles (\textit{kreise}) in Graub\"unden also
provided religious autonomy [12]. The intentional formation of
political boundaries in regions that would have violence according to
the model, and the subsequent model propensity below the threshold
associated to a lack of actual violence are consistent with the
hypothesis on the role of boundaries in peaceful coexistence.

\bigskip

\begin{figure}
\includegraphics[width=14cm]{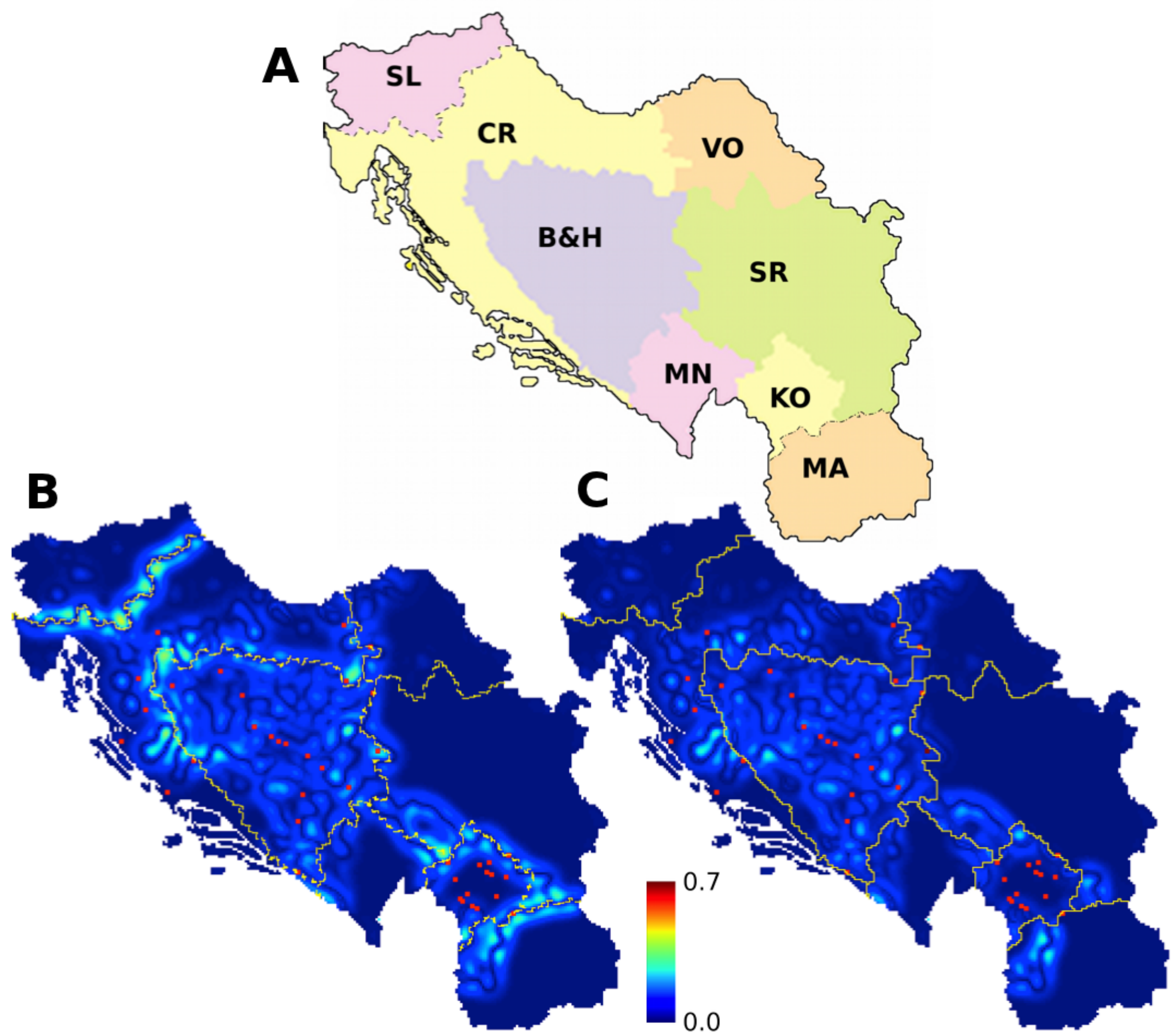}
\caption{\label{fig3} {\bf (A)} Map of the area of the former
  Yugoslavia showing administrative provinces.  Propensity to violence
  calculated without {\bf (B)} and with {\bf (C)} administrative
  boundaries, using a characteristic length of 21\,km. Locations of
  boundaries are shown on both plots as solid and dashed yellow lines
  respectively. Sites of reported violence are shown as red dots
  [18]. Spurious violence is predicted along the borders of Slovenia
  and Macedonia when boundaries are not included. Province labels are:
  SL: Slovenia, CR: Croatia, VO: Vojvodina*, B\&H: Bosnia \&
  Herzegovina, SR: Serbia, MN: Montenegro, KO: Kosovo*, MA:
  Macedonia. (*Autonomous administrative provinces of Serbia.)}
\end{figure}

\noindent\textit{Yugoslavia} - Our modified method including
boundaries was tested on the previous case study of Yugoslavia,
consisting of the combined area of Croatia, Bosnia, Serbia and
Montenegro.  Topographical boundaries reduce the maximum propensity
from 0.63 to 0.57, and administrative borders to 0.56. The
correlations of predicted and reported violence changes were
insignificantly lower, with correlations of 0.86 and 0.85,
respectively. That political boundaries do not have a greater impact
on the calculated violence implies that they do not align with the
geographical boundaries between groups. We also extended the area to
include Macedonia and Slovenia, parts of the Socialist Federal
Republic of Yugoslavia before gaining independence
(Fig. \ref{fig3}). With the political boundaries the correlation is
still 0.85; however, when political boundaries are not included, the
correlation is reduced considerably to 0.72. The lower correlation is
specifically due to a high calculated propensity to violence along the
borders of Slovenia with Croatia, and of Macedonia with Serbia and
Kosovo. These areas, however, were peaceful---consistent with the
predictions when boundaries are included. Our results suggest that
these political borders were instrumental in reducing ethnic violence,
whereas the violence in other areas of Yugoslavia was not prevented
because of poor alignment of borders with population groups.

This work is part of a broader effort to use new methods for
quantitative analysis of patterns of violence and their prevention
[23--31]. There is also interest in ethnic group interactions across 
national borders [32-34].
We have shown that groups that are not well-mixed but are
geographically separated by natural or political boundaries into
autonomous domains are peaceful in both Switzerland and the former
Yugoslavia. Our work clarifies the ambiguities of mixed language and
religion Swiss cantons by showing that in most cases the natural
geography of the populations conspires to lead to a low level of
violence, so that additional boundaries were not necessary; where they
were needed, as in Graub\"unden, they were established. The highest
calculated propensity to violence is between linguistic groups in the
northern part of the canton of Bern, where historically unresolved
real world tensions actually exist. Our analysis indicates that both
administrative and natural barriers can play a significant role in
mitigating conflict between religious and linguistic
groups. Historical evidence suggests that for religious groups the
boundaries in Switzerland were created to provide autonomy to a group
with a shared identity and avoid conflict among multiple
groups. Ongoing efforts to reduce tensions in Bern include introducing
new political boundaries. The many political, social and economic
factors that play roles in reducing violence [4--6,9--14] build on a
strong foundation of geographical borders. Our analysis suggests that
when partition within a country is viewed as an acceptable form of
conflict mitigation, such partition can give rise to highly stable
coexistence and peace.

We thank Stuart Pimm, Irving Epstein and Lawrence Susskind for helpful comments on the
manuscript, Michael Widener and Blake Stacey for help with a figure 
and formatting. This work was supported in part by AFOSR under 
grant FA9550-09-1-0324 and ONR under grant N000140910516.

\bigskip

\noindent \textbf{References:}

[1] M. Lim, R. Metzler, Y. Bar-Yam, Global pattern formation and
ethnic/cultural violence.  Science 317, 1540 (2007).

[2] Imagine Coexistence: Restoring humanity after violent ethnic conflict, 
A. Chayes, M. L. Minow eds., (Jossey-Bass, San Francisco, 2003) 

[3] C. Kaufmann, When all else fails: Ethnic population
transfers and partitions in the Twentieth Century. Int. Secur. 23, 120
(1998).

[4] A. Lijphart, Democracy in Plural Societies: A Comparative
Exploration (Yale University Press, New Haven, 1977).

[5] C. L. Schmid, Conflict and Consensus in Switzerland (University of
California Press, Berkeley, 1981).

[6] W. Martin, A History of Switzerland: An Essay on the Formation of
a Confederation of States (G. Richards, London, 1931).

[7] U. Im Hof, Geschichte der Schweiz, vol. 188 of Kohlhammer
Urban-Taschenb\"ucher (Kohlhammer, Stuttgart, 1991).

[8] N. Bonaparte, Proclamation de St. Cloud, 30 septembre 1802 in
Bonaparte et la Suisse: Travaux Preparatoires de l'Acte de Mediation
(1803), V. Monnier, Ed. (Helbing \& Lichtenhahn, Geneva, 2002).

[9] J. Steiner, Amicable Agreement Versus Majority Rule: Conflict
Resolution in Switzerland (University of North Carolina Press, Chapel
Hill, 1974).

[10] H. E. Glass, Ethnic diversity, elite accommodation and federalism
in Switzerland. Publius 7, 31 (1977).

[11] W. Linder, Swiss Democracy: Possible Solutions to Conflict in
Multicultural Societies (Palgrave Macmillan, New York, ed. 3, 2010).

[12] R. C. Head, Early Modern Democracy in the Grisons: Social Order
and Political Language in a Swiss Mountain Canton: 1470--1620
(Cambridge University Press, Cambridge, 2002).

[13] K. D. McRae, Conflict and Compromise in Multilingual Societies:
Switzerland (Wilfrid Laurier University Press, Waterloo, Ontario,
1983).

[14] C. H. Church, The Politics and Government of Switzerland
(Palgrave MacMillian, New York, 2004).

[15] T. Christin, S. Hug, Federalism, the geographic location of
groups, and conflict, CIS Working Paper No. 23, Center for Comparative
and International Studies, ETH Zurich and University of Zurich (2006).

[16] W. Shearer, Determine Indicators for Conflict Avoidance, Science
(2008)

[17] P. Ch. Ivanov et al., Nature 383, 323 (1996).

[18] I. Daubechies, Ten Lectures on Wavelets, (SIAM, Philadelphia,
1992).

[19] A. Jarvis, H. I. Reuter, A. Nelson, E. Guevara, Hole-filled SRTM
Version 4: \url{http://srtm.csi.cgiar.org} (2008).

[20] A. Bovik, Essential Guide to Image Processing (Academic Press,
Burlington, 2009).

[21] W. R. Keech, Linguistic diversity and political conflict: Some
observations based on four Swiss cantons. Comp. Politics 4, 387
(1972).

[22] ``Citizens to settle territorial Jura conflict,'' Swissinfo, May
4, 2009
(\url{http://www.swissinfo.ch/eng/politics/Citizens_to_settle_territorial_Jura_conflict.html?cid=7377228},
Accessed Feb. 21, 2011).

[23] D. L. Horowitz, Ethnic Groups in Conflict (University of
California Press, Berkeley, ed. 2, 2000).

[24] B. Harff, T. R. Gurr, Ethnic Conflict in World Politics
(Westview, Boulder, ed. 2, 2004).

[25] M. Reynal-Querol, Ethnicity, Political Systems, and Civil
Wars. J. Conflict Res. 46, 29 (2002).

[26] J. Fox, Religion, Civilization, and Civil War: 1945 Through the
New Millennium (Lexington Books, Lanham, MD, 2004).

[27] I. S. Lustick, D. Miodownik, R. J. Eidelson, Secessionism in
multicultural states: Does sharing power prevent or encourage it?
Am. Pol. Sci. Rev. 98, 209 (2004).

[28] T. R. Gulden, Spatial and temporal patterns in civil violence:
Guatemala, 1977-1986.  Politics Life Sciences 21, 26 (2002).

[29] L. E. Cederman, K. S. Gleditsch, Introduction to special issue on
``Disaggregating Civil War.'' J. Conflict Res. 53, 487 (2009).

[30] H. Buhaug, S. Gates, The geography of civil war. J. Peace
Res. 39, 417 (2002).

[31] J. C. Bohorquez, S. Gourley, A. R. Dixon, M. Spagat,
N. F. Johnson, Common ecology quantifies human insurgency. Nature 462,
911 (2009).

[32] African Boundaries: Barriers, Conduits and Opportunities, 
P. Nugent, A. I. Asiwaju, eds. (Pinter, London, 1996).

[33] K. Mitchell, Transnational discourse: bringing geography back in, 
Antipode 29, 101 (1997).

[34] M. Silberfein, A. Conteh, Boundaries and conflict in the Mano River 
region of West Africa, Conflict Management and Peace Science 23, 
343 (2006).

\newpage
\counterwithin{figure}{section}
\noindent {\large Appendices for:
Good Fences: \\ The Importance of Setting Boundaries for Peaceful Coexistence}

\appendix

\tableofcontents
\setcounter{tocdepth}{2}

\section{Methods}

\subsection{Identifying the propensity to violence using a wavelet filter}
\label{wavelet}
The potential for conflict is quantified in our model using a wavelet
filter [A.1--A.3]. In essence, the filter evaluates the extent of
the presence of a type in a circular area with a specified radius and
subtracts from this the presence of the same type in a surrounding
area. This results in cancellation if the same type is located in the
surrounding area. Other types are all treated with the opposite sign
causing cancellation if there are mixed populations of the first type
with the others. Thus, the largest values are obtained for an island
of one type surrounded by other types.  Large values are also obtained
for a peninsula of one type into a sea of other types. To evaluate the
likelihood of violence at a particular location, we apply the filter,
centered at that location, for each of the types. The likelihood of
violence in that region is the maximum over all types. Unlike the
earlier method [A.4], we included all population types on each site
of a grid rather than basing calculations on an agent
model. Mathematically the expression for the filter applied at a
location $(x,y)$, with the maximum taken over all types, is
\begin{equation}
c(x,y) = \max_s \sum_{x',y'} m(x-x',y-y')
  \left(p_s(x',y') - \sum_{s' \neq s} p_{s'}(x',y') \right),
\end{equation}
which is a convolution of the fraction of the population of one type,
$p_s(x,y)$ minus the fractional population of other types, with a
wavelet,
\begin{equation}
m(x,y) = (1 - \rho(x,y)^2) e^{-\rho(x,y)^2},
\end{equation}
where the scaled distance from the center is given by 
\begin{equation}
\rho(x,y) = \frac{\sqrt{x^2 + y^2}}{r_c}, 
\end{equation}
the Euclidean distance divided by the radius of the wavelet, $r_c$,
which is half of the diameter, $l_c$, the model parameter identifying
the size of groups that are likely to engage in conflict. The value of
$c(x,y)$ serves as a measure of the likelihood of violence in the
vicinity of the location $(x,y)$.  When performing statistical tests
on the prediction of violence, we specify a threshold that
distinguishes regions of violence from regions of non-violence
according to whether $c(x,y)$ exceeds the specified threshold.

\subsection{Boundaries}
\label{boundaries}
We model both topographical and administrative boundaries within a
country as preventing intergroup violence across them, similar to
national boundaries in the earlier method [A.4]. A cliff separating a
plateau from a plain is considered to be a barrier to movement between
the upper and lower areas and thus serves as a boundary.

We generalize the previous method for incorporating boundaries to
allow for partial boundaries and boundaries with gaps. Partial
boundaries between areas within the country can arise due to
mountains, lakes, or at convoluted political borders. For such
boundaries, we consider the line of sight from a given location to
identify the populations which impact on the propensity for violence
at that location. Populations outside of the line of sight are not
included as contributing to violence. Thus an effective map of
populations as experienced at each site is constructed, determined by
the specific orientation of any boundaries relative to that site. The
areas which are blocked from sight are populated with a neutral
population, the existing local proportions of the population. This
better matches both the mixed and single type local populations than a
single type. The local proportions were measured within a range of two
characteristic lengths (wavelet diameters) of each site, considering
only sites that are in a line of sight.

\subsection{Empty sites}
\label{empty sites}
Some small areas are unpopulated. These and lake areas were treated as
other sites, but the violence at these sites was set to zero. Only
small differences arise if these unpopulated areas are treated
differently.

\begin{figure}
\includegraphics[width=14cm]{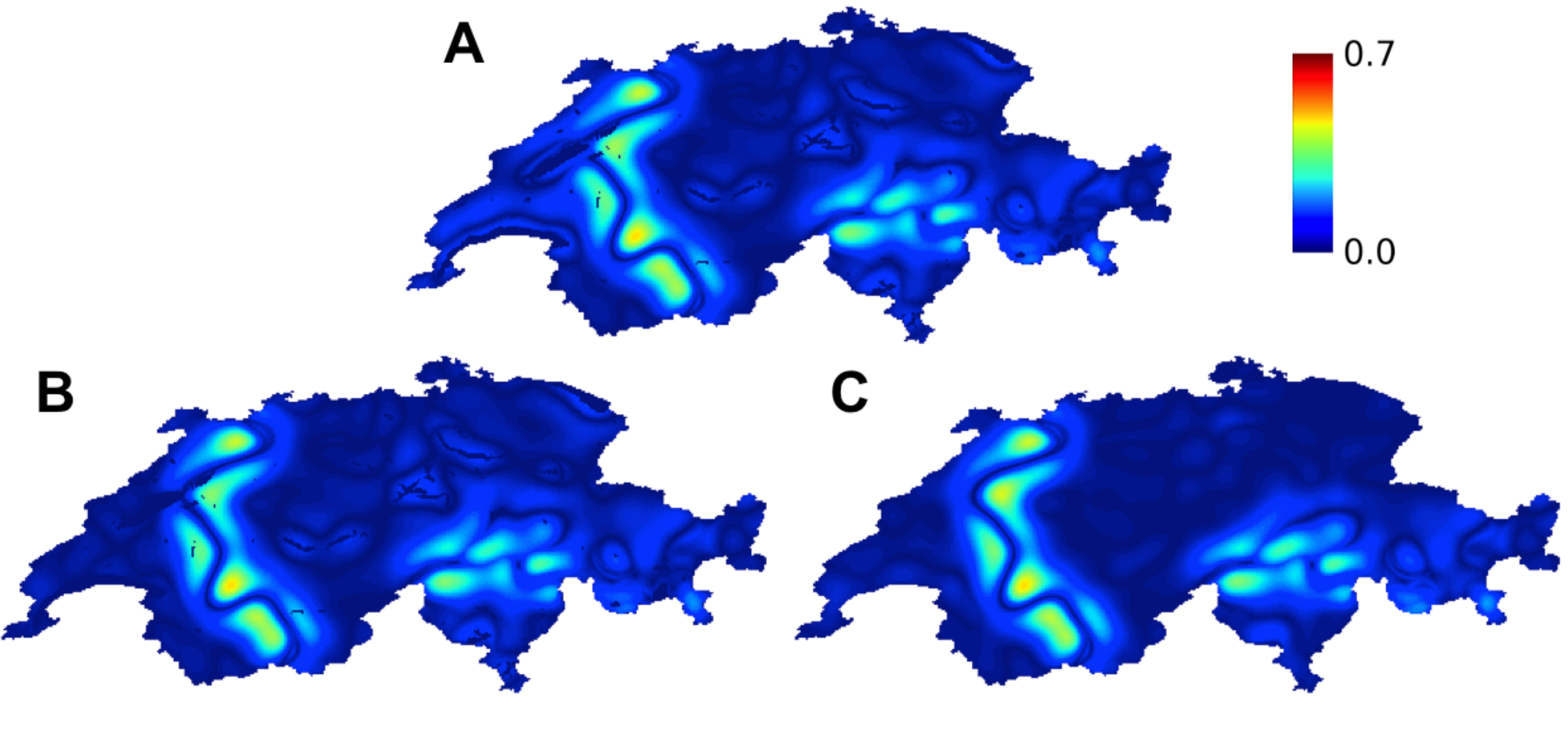}
\caption{\label{figS1} Level of predicted violence between linguistic
  groups in Switzerland using a characteristic length scale of
  24\,km. Each panel represents results for a different treatment of
  lakes and unpopulated land areas: {\bf (A)} including lakes and
  unpopulated land areas as empty sites; {\bf (B)} including as
  barriers the lakes of Leman and Neuchatel; {\bf (C)} interpolating a
  composition for all unpopulated sites from neighboring sites.}
\end{figure}

There are two types of unpopulated areas, land and water. Unpopulated
land areas are treated as other land areas for the purpose of the
calculations. After the calculation we set the propensity to violence
in those locations to zero. The results were not affected
significantly (Fig. \ref{figS1}). Water areas were treated similarly,
with the exception that bodies of water that are large were considered
to be topographical barriers, similar to mountains and
cliffs. Specifically, we included the two largest lakes, Leman and
Neuchatel, both of which have a length above 10\,km, which is
comparable to the range of characteristic length scales used to detect
a propensity to violence.\\

\noindent \textbf{References:}

[A.1] I. Daubechies, Ten Lectures on Wavelets, (SIAM, Philadelphia, 1992).

[A.2] A. Arneodo, E. Bacry, P.V. Graves, J.F. Muzy,
Phys. Rev. Lett. 74, 3293 (1995).

[A.3] P. Ch. Ivanov, M. G. Rosenblum, C-K. Peng, J. Mietus,
S. Havlin, H.E. Stanley, A.L.  Goldberger, Nature 383, 323 (1996)

[A.4] M. Lim, R. Metzler, Y. Bar-Yam, Global pattern formation and
ethnic/cultural violence.  Science 317, 1540 (2007).

\section{Census data}

The commune composition used in our calculations was based on the
census of 2000 and 1990 published by the Swiss Statistical
Office. Where municipalities have merged, an aggregate of their
previous constituent municipalities was taken. Three official
languages we considered are French, German and Italian, which comprise
91\% of the total population. The fourth official language, Romansch,
is 2\%. The religions considered are Roman Catholic and Protestant
accounting for 77\% of the total with less than 8\% belonging to other
religious groups and the remainder not subscribing to a religion or
not specifying one. The 1990 census data is only readily available on
a cantonal level. As described in Section \ref{religion 1990}, we
estimated the commune composition using the 2000 value and the change
in the parent canton between 1990 and 2000.

\section{Summary of model comparisons with the data}
\label{model comparisons}
We briefly summarize the comparisons between model predictions and the
observed data reported in the main section of the paper.

Our examination of linguistic and religious groups in Switzerland
highlighted cases where violence is predicted without the presence of
boundaries, but is mitigated by the consideration of topographical and
political boundaries appropriate to linguistic and religious groups,
respectively.

(1) Topographical boundaries reduced violence between linguistic
groups. This occurred along (a) Alpine boundaries of the Swiss Alps
between German-speaking and Italian-speaking populations, (b) Alpine
boundaries between German-speaking and French-speaking populations,
and (c) Jura range boundaries between German-speaking and
French-speaking populations.

(2) Political boundaries reduced violence between religious
groups. This is the case both for (a) canton boundaries and for (b)
circle boundaries in the canton of Graub\"unden.

Our analysis also identified locations in which our model does not
predict violence despite linguistic or religious heterogeneity and no
explicit boundaries.

(3) The straightness of the boundary prevents violence between
linguistic groups in Fribourg/Freiburg.

(4) Isolation of a Protestant population on an appendage from the
Catholic majority prevents violence in Fribourg/Freiburg.

We also identified one area at the highest level of calculated
residual propensity to violence and it corresponds to an area of
unresolved historical conflict.

(5) The northeastern part of the canton of Bern is the location of
both the highest prediction of propensity to violence, and a
real-world history of intergroup tension. The unique condition of the
conflict in this part of Switzerland and its correspondence to the
prediction by the model provides additional confirmation of the model.

Considering the predicted and reported violence in the former
Yugoslavia also demonstrated the importance of the boundaries which
coincide with ethnic divisions.

(6) Political boundaries between Slovenia and Macedonia and the other
countries of the former Yugoslavia prevent violence along their
borders.

(7) The borders between the countries of Croatia, Bosnia, Serbia and
Montenegro were not aligned with the boundaries between ethnic groups
and so were ineffective at reducing violence.

\section{Languages}
\label{languages}
\begin{figure}
\includegraphics[width=14cm]{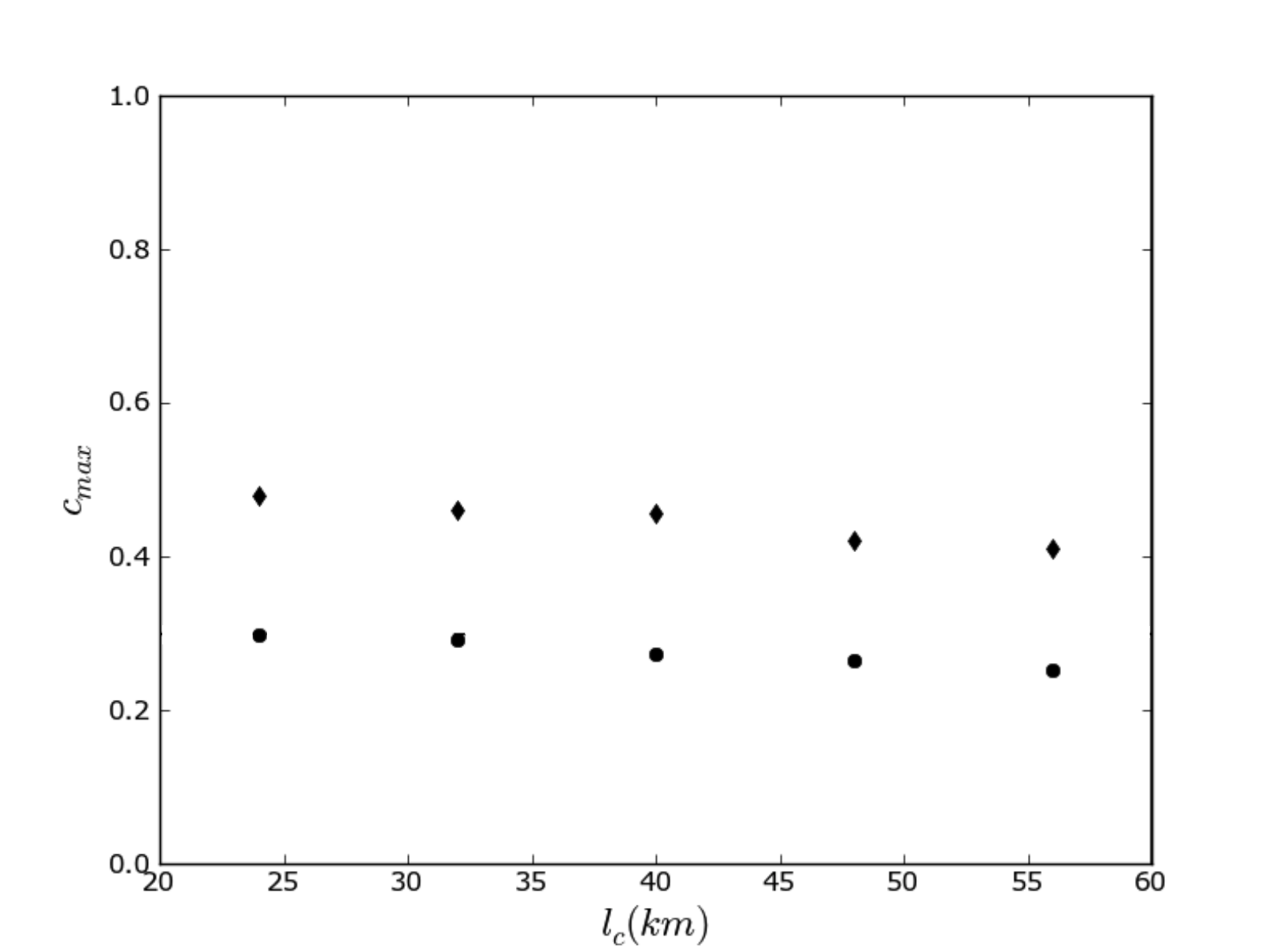}
\caption{\label{figS2} Maximum level of the propensity to violence
  between linguistic groups in Switzerland as calculated in the model
  as a function of the characteristic length scale. The calculation is
  performed with effect of topographical boundaries ($\bullet$) and
  without effect of topographical boundaries ($\Diamond$).}
\end{figure}

\begin{figure}[ht]
\includegraphics[width=14cm]{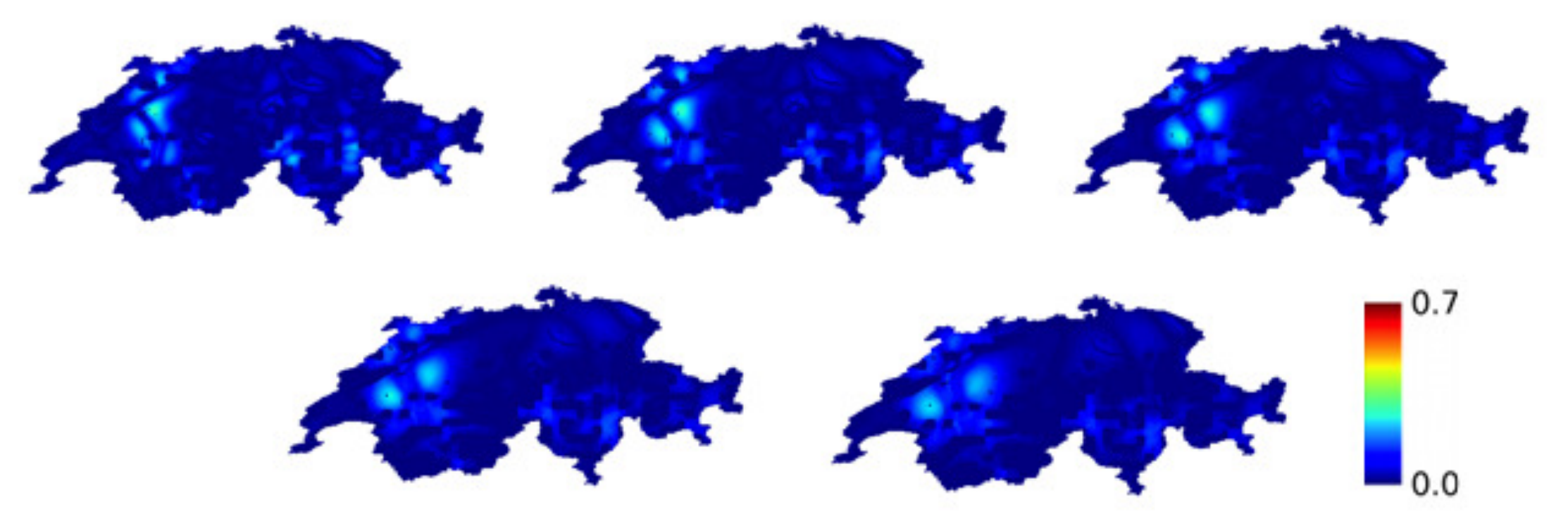}
\caption{\label{figS3} Level of propensity to violence between
  linguistic groups in Switzerland including the effect of
  topographical boundaries. Characteristic lengths increases from left
  to right, top to bottom with the values 24, 32, 40, 48, 56\,km.}
\end{figure}

\begin{figure}[ht]
\includegraphics[width=14cm]{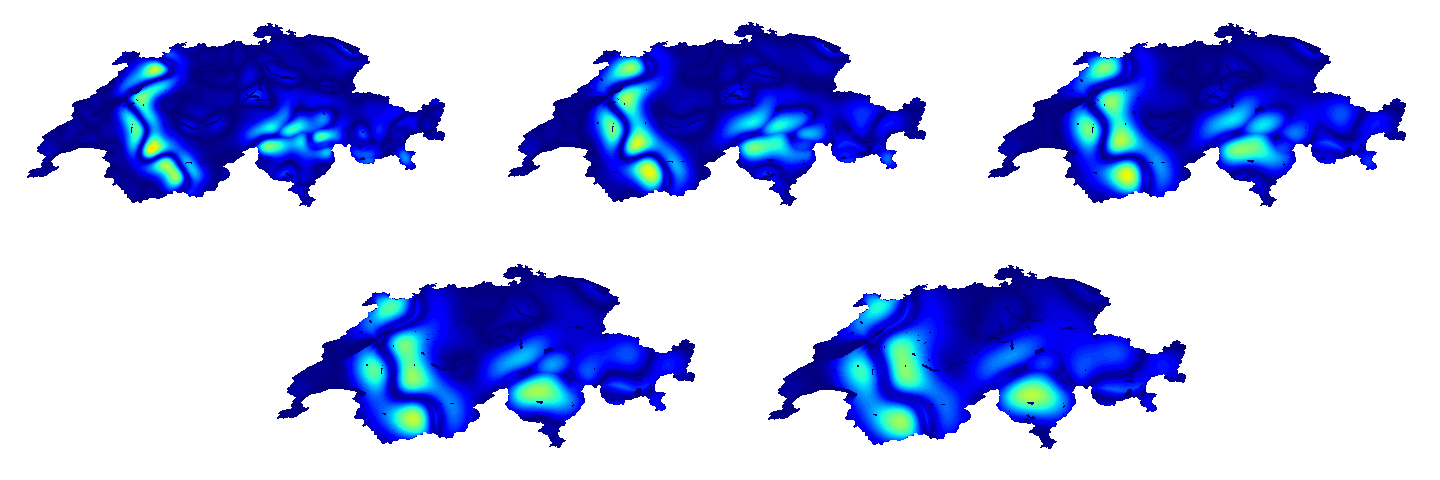}
\caption{\label{figS4} As in Fig. \ref{figS3} without the effect of
  topographical boundaries.}
\end{figure}

Here we describe in greater detail the results of the calculation of
the propensity to violence between linguistic groups in Switzerland
with and without the effect of topographical boundaries. In the main
text we described the calculation of propensity for violence for a
characteristic length scale of 24\,km. Here we provide it for the
length scales 24, 32, 40, 48 and 56\,km. Figs. \ref{figS2}--\ref{figS4}
show that, at all values of the characteristic length scale, the
propensity for violence is high for calculations without topographical
boundaries and is dramatically reduced by their inclusion.

\section{Elevation edges}
\label{elevation edges}
\begin{figure}
\includegraphics[width=14cm]{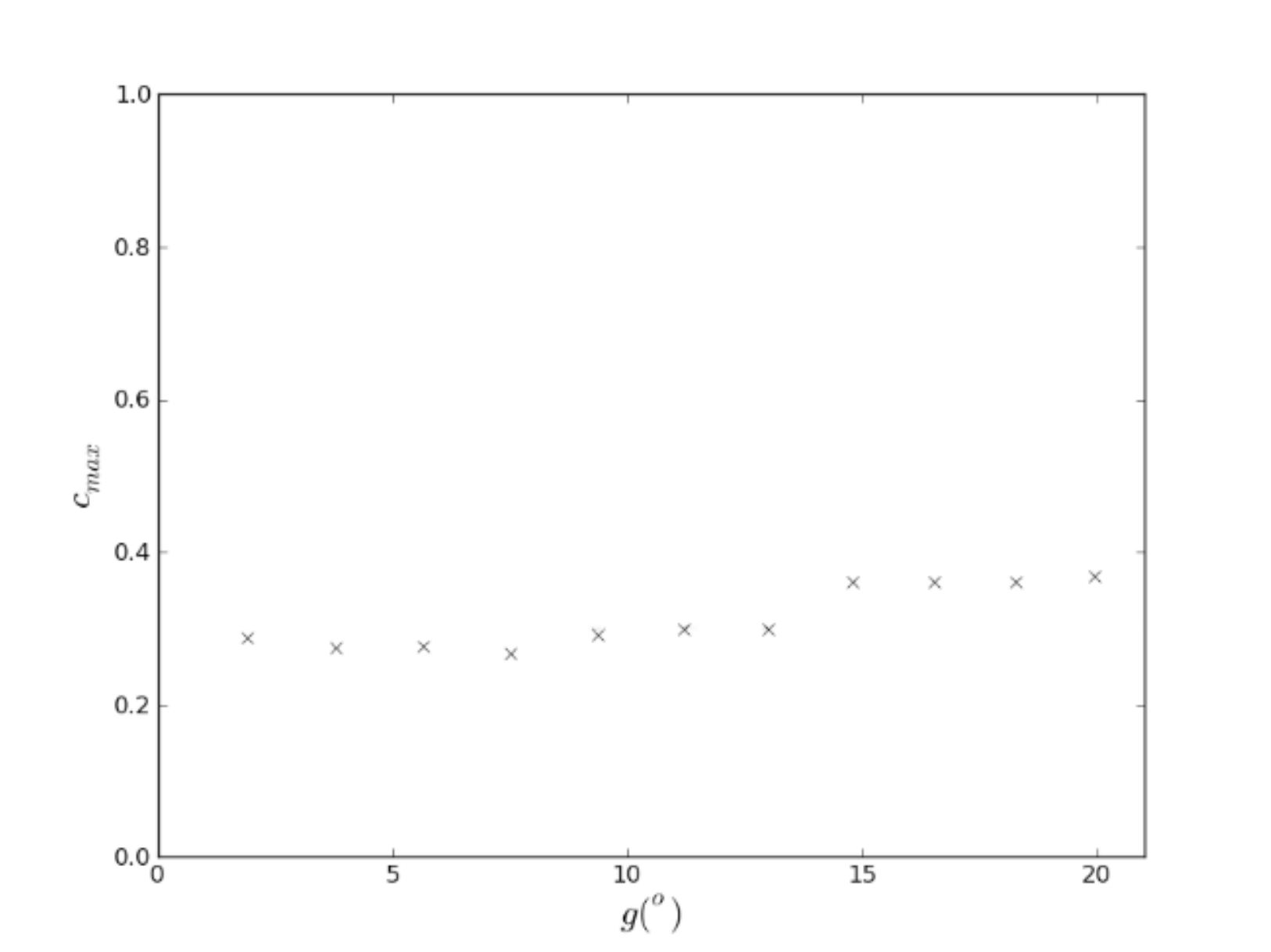}
\caption{\label{figS5} The maximum propensity to violence between
  linguistic groups in Switzerland as the threshold gradient for
  topographical barriers varies.}
\end{figure}

\begin{figure}
\includegraphics[width=14cm]{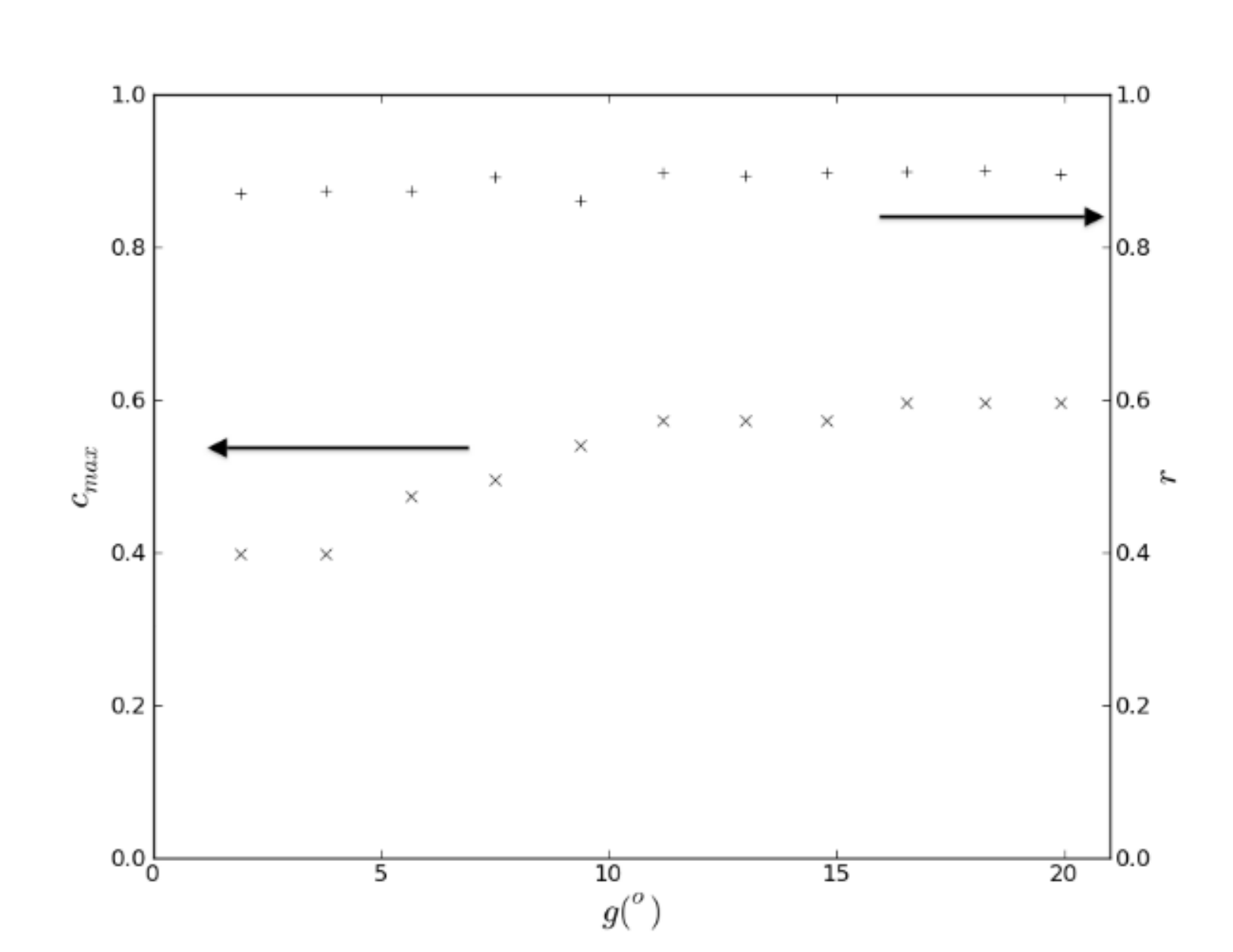}
\caption{\label{figS6} Maximum propensity to violence ($\times$, left
  axis) and correlation of predicted with reported violence (+, right
  axis) in the former Yugoslavia as the threshold gradient for
  topographical barriers $g$ varies.}
\end{figure}

Here we investigate the robustness of our analysis to variation of the
calculation of topographical barriers extracted from the elevation
data. We vary the gradient threshold that determines the presence of a
boundary and compare the results for linguistic groups in
Switzerland. We also include here a similar comparison of the
calculation of the impact of topographical edges on the conflict
between ethnic groups within the former Yugoslavia.  Figure
\ref{figS5} shows the variation of the maximum propensity to violence
in Switzerland as the threshold gradient for geographical barriers
varies. The propensity is robust to the variation across a range of
angles. Still, as the gradient increases and barriers are removed the
propensity to violence increases. The model results are consistent
with the expectation that it is necessary to include geographical
features as barriers in order to achieve agreement with the locations
of actual reports of violence, and is consistent with the hypothesis
that such barriers are effective in mitigating outbreaks of violence.
Figure \ref{figS6} shows the maximum propensity to violence calculated
for the former Yugoslavia as a function of changes in the gradient
threshold, and the resulting correlation of predicted and reported
violence. The results show that while some variation in the maximum
value of the predicted violence propensity occurs, it remains above
the threshold for expected violence. The correlation with observed
violence is not very sensitive to the gradient of the edges in
elevation.  This indicates that areas of predicted violence continue
to be proximate to the areas of reported violence. Topographical
features are not sufficiently steep or aligned with the boundaries of
population groups to inhibit violence.

\section{Religion (2000 census)}
\label{religion 2000}
Here we describe in greater detail the calculation of violence between
religious groups in Switzerland. In the main text we described the
calculation of propensity for violence for a characteristic length
scale of 24\,km. Here we provide the results for the length scales 24,
32, 40, 48 and 56\,km.

\begin{figure}
\includegraphics[width=14cm]{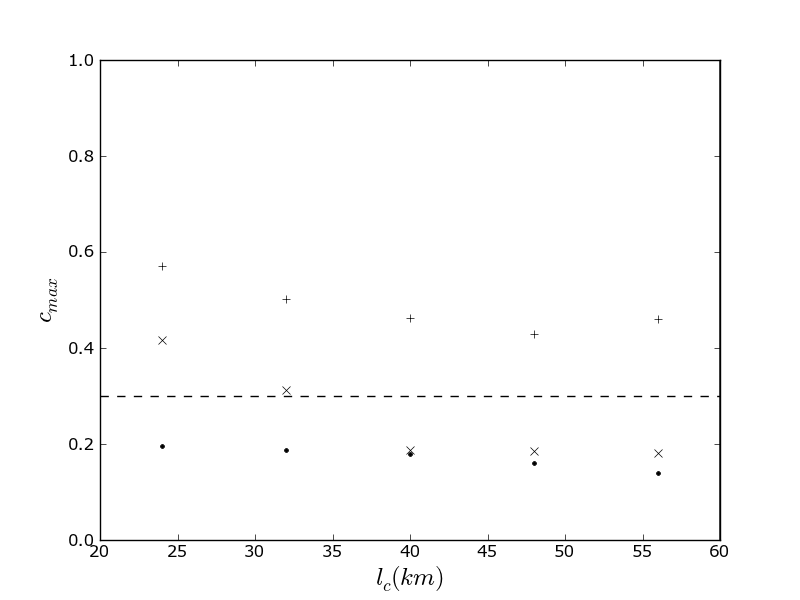}
\caption{\label{figS7} Maximum level of the propensity to violence
  between religious groups in Switzerland as a function of
  characteristic length scale according to the model. Calculations are
  shown including the effect of canton boundaries and Graub\"unden
  circle boundaries ($\bullet$), including the effect of canton
  boundaries only ($\times$), and without the effect of political
  boundaries (+). The dashed line represents the inferred threshold of
  propensity of violence in order for violence to occur.}
\end{figure}

\begin{figure}[t]
\includegraphics[width=13.5cm]{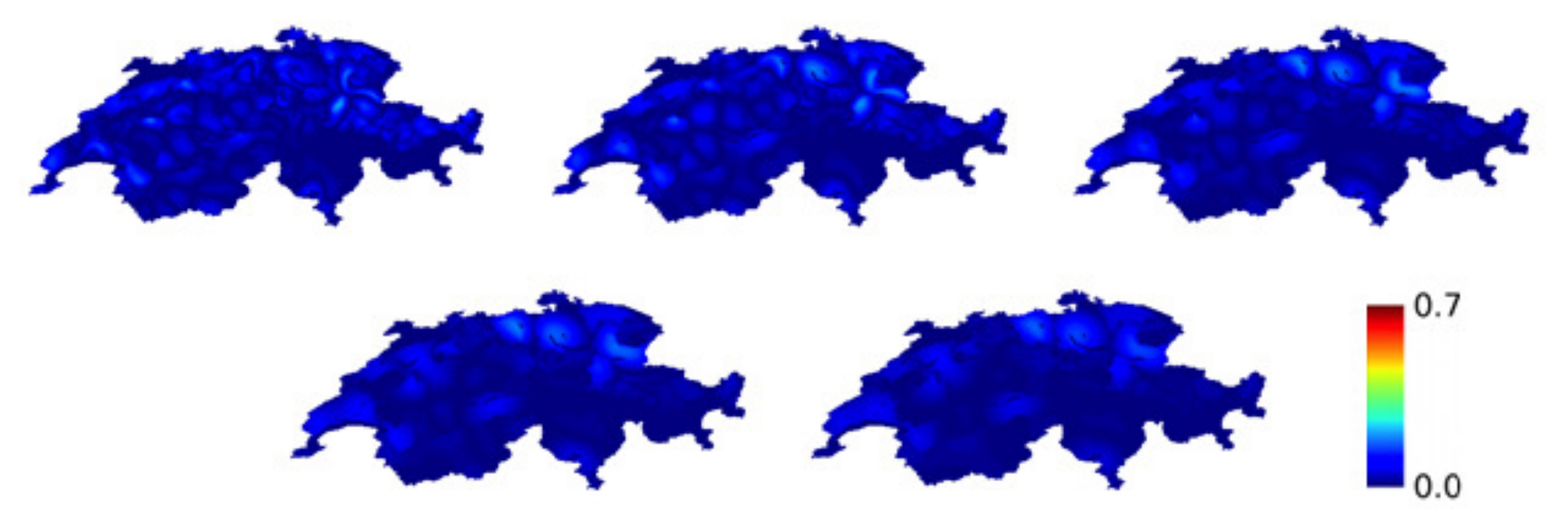}
\caption{\label{figS8} Level of predicted violence between religious
  groups in Switzerland with political boundaries, including both
  cantons and Graub\"unden circles (2000 census). Characteristic
  length increases from left to right, top to bottom for the values
  24, 32, 40, 48, 56\,km.}
\end{figure}

\begin{figure}[t]
\includegraphics[width=13.5cm]{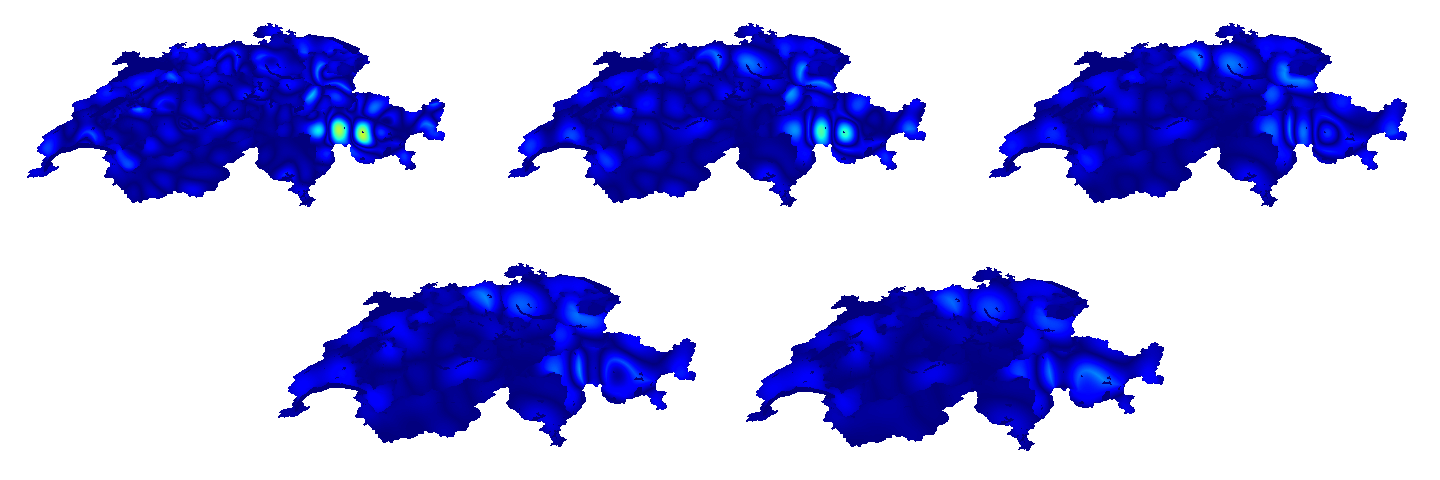}
\caption{\label{figS9} As in Fig. \ref{figS8} but including only the
  effect of canton boundaries.}
\end{figure}

\begin{figure}[b]
\includegraphics[width=13.5cm]{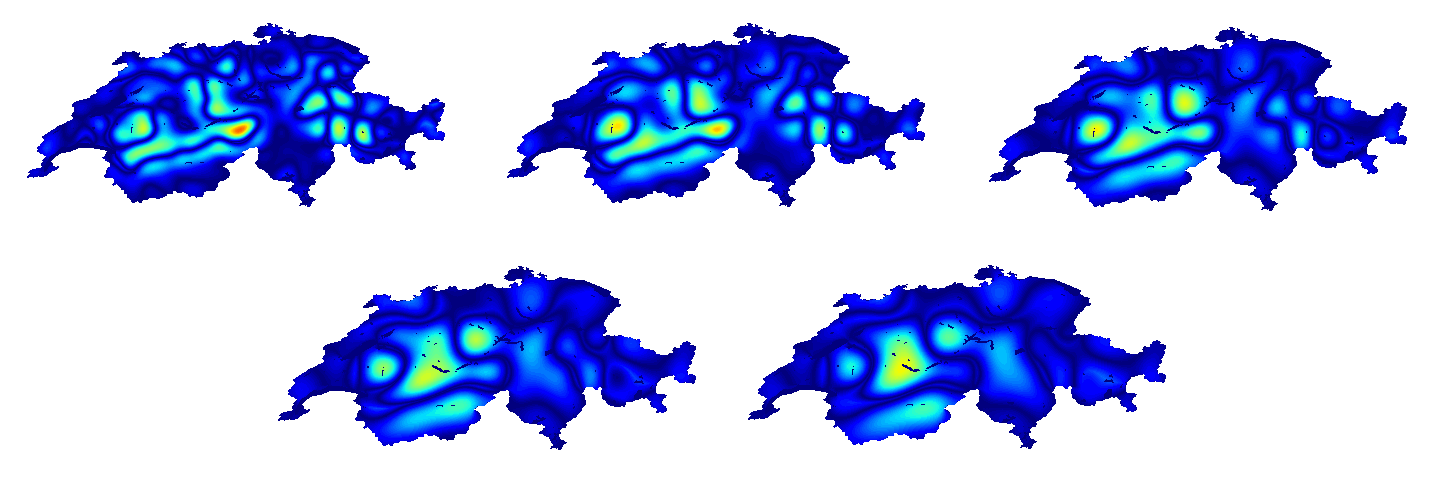}
\caption{\label{figS10} As in Fig. \ref{figS8} but without the effects
  of political boundaries.}
\end{figure}

Figure \ref{figS7} plots the maximum propensity to violence with
canton and Graub\"unden circle boundaries, with canton boundaries
only, and without political boundaries. The corresponding maps are
shown in Figs. \ref{figS8}--\ref{figS10}. Autonomy within cantons and
Graub\"unden circles has been established to prevent
conflict. Consistent with the historical experience, the model results
imply that without these boundaries violence would be expected, but
with them it is not. The effect of canton boundaries is important
across all length scales, that of the circles in Graub\"unden is
important at the smaller length scales. This result specifically
suggests that length scales of 24--32\,km correspond to a geographical
group size that is susceptible to violence.

\clearpage

\section{Religion (1990 census)}
\label{religion 1990}
In the main paper we reported the propensity for violence between
religious groups for the 2000 census for the characteristic length of
24\,km. During the 1990s there was a significant reduction in religious
affiliation. We therefore considered also the 1990 census. The results
are very similar to those of the 2000 census with maximum propensity
without boundaries of 0.59 (compared to 0.57) reduced to 0.23 when
including the political boundaries (compared to 0.20).

\begin{figure}[ht]
\includegraphics[width=11cm]{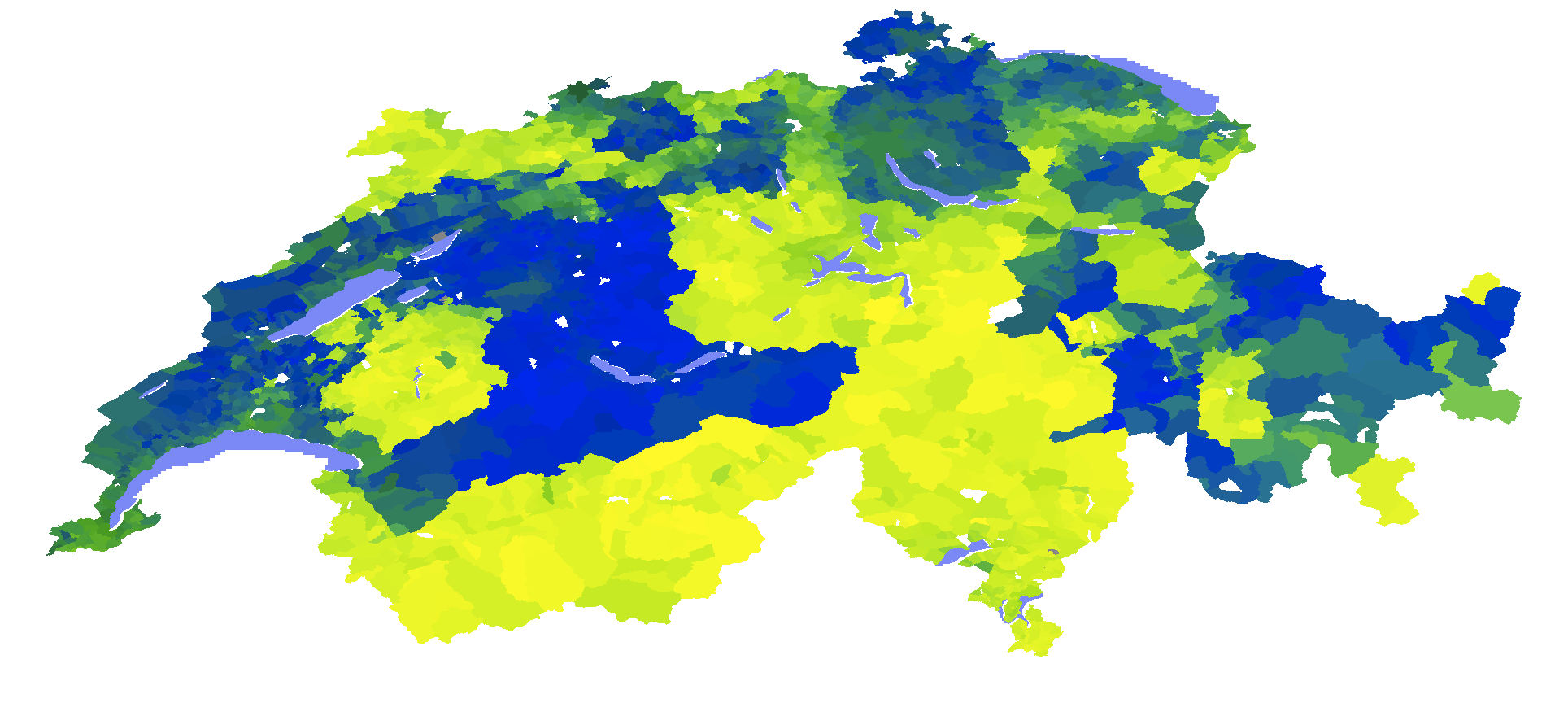}
\includegraphics[width=2cm]{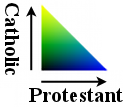}
\caption{\label{figS11} Proportion of religious groups according to
  interpolated 1990 census. Communes are colored according to
  proportion of Protestant (blue) and Catholic (yellow) as shown by
  color triangle.}
\end{figure}

\begin{figure}[ht]
\includegraphics[width=13.5cm]{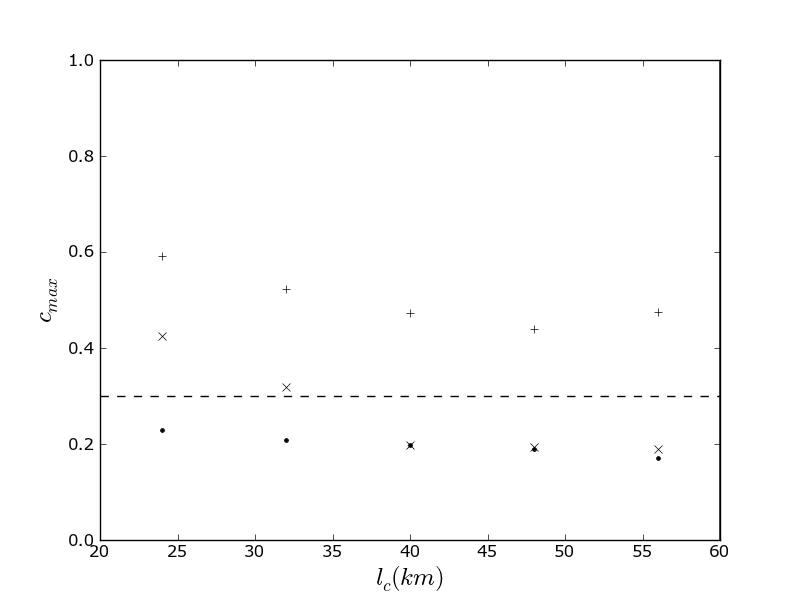}
\caption{\label{figS12} As in Fig. \ref{figS7} for the 1990 census.}
\end{figure}

\begin{figure}[ht]
\includegraphics[width=13.5cm]{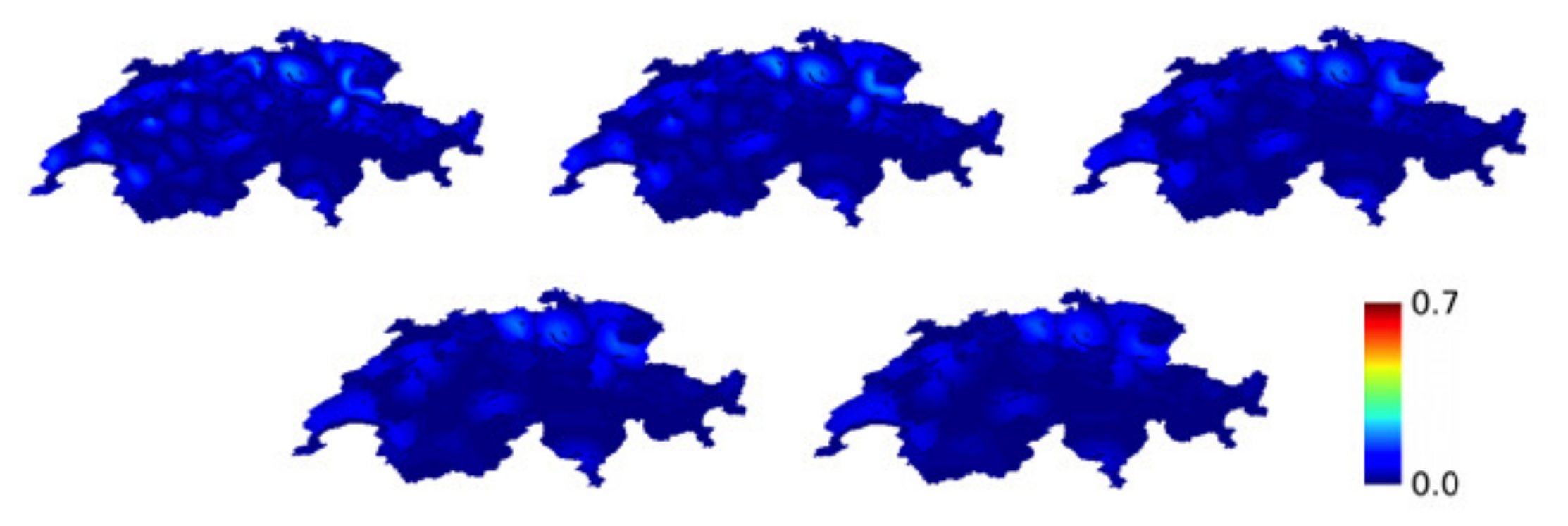}
\caption{\label{figS13} As in Fig. \ref{figS8} for the 1990 census.}
\end{figure}

\begin{figure}[ht]
\includegraphics[width=13.5cm]{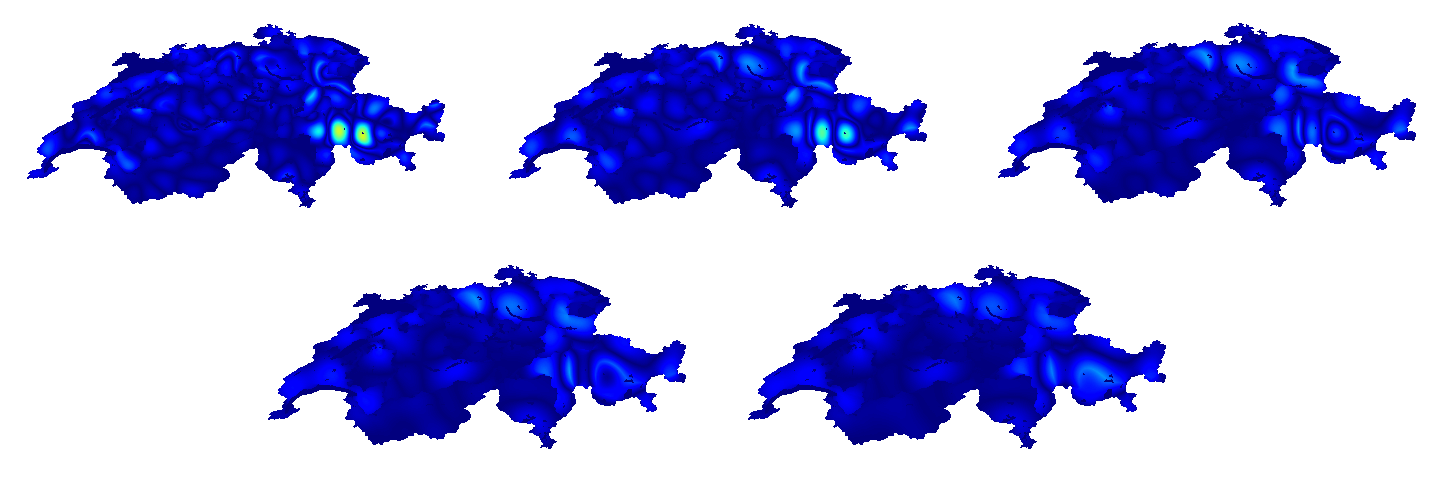}
\caption{\label{figS14} As in Fig. \ref{figS9} for the 1990 census.}
\end{figure}

\begin{figure}[ht]
\includegraphics[width=13.5cm]{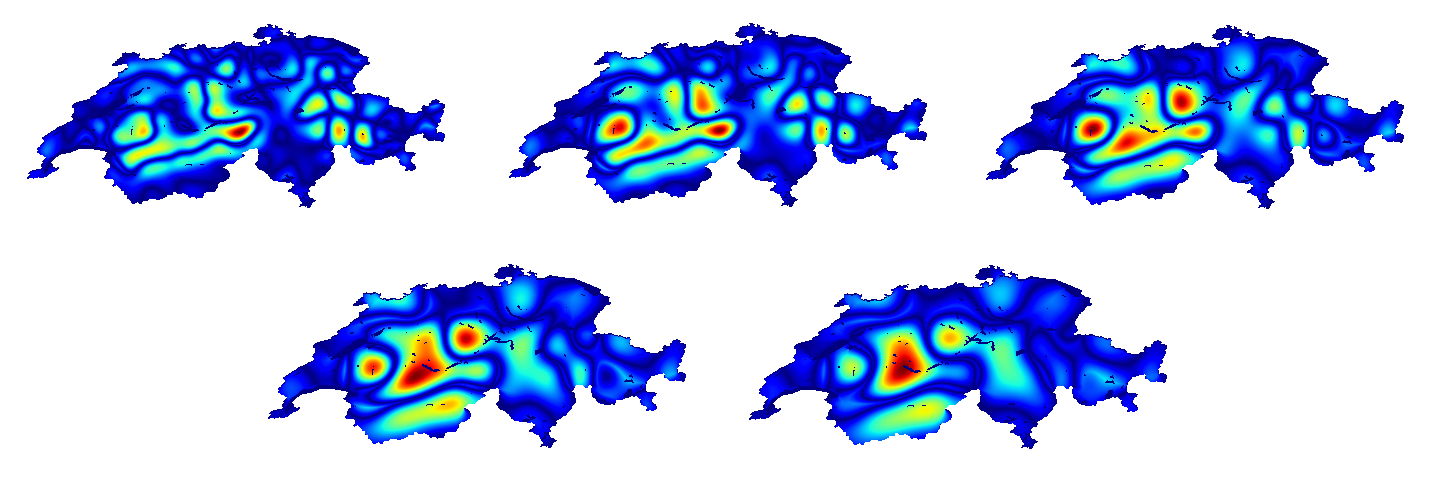}
\caption{\label{figS15} As in Fig. \ref{figS10} for the 1990 census.}
\end{figure}

In 2000 Roman Catholicism and Protestantism accounted for 87\% of the
population, 10\% more than in 2000, and with only 9.5\% identifying
themselves as atheist or not specifying religious affiliation. The
census for religions in Switzerland in 1990 is readily available only
at a canton level resolution rather than the municipality level used
in our calculations. We used the reduction of religious affiliation in
the entire canton to estimate religious composition for each
municipality in 1990. Explicitly:
\begin{equation}
p = p' \times \frac{1.0 - \beta a'}{1.0 - a'},
\end{equation}
where $p$ and $p'$ are the value of the municipal Catholic or
Protestant proportion of the population estimated for 1990 and given
for 2000, $a'$ is the unaffiliated municipality population proportion
in 2000, and 
\begin{equation}
\beta = \frac{A}{A'}
\end{equation}
is the ratio of unaffiliated canton population proportions, $A$ and
$A'$, in 1990 and 2000. Fig. \ref{figS11} is a map of the resulting
religious affiliation. Figs. \ref{figS12}--\ref{figS15} show the
calculations of the propensity for violence for the 1990 census
corresponding to the results for the 2000 census results shown in
Fig. \ref{figS7}--\ref{figS10}.

\clearpage

\clearpage

\clearpage

\section{Bern/Jura violence}
Unique in Switzerland in recent decades, the violence in the area of
Bern/Jura based on linguistic conflict included targeted arson and
bombings and a violent encounter between demonstrators.  We performed
an analysis of the correlation of reported violence with the location
of highest propensity calculated by the theory, which is reduced by
local geography compared to what would be expected without it. The
resulting correlation is greater than 0.95.  We note that the
difficulty in relieving the conflict in the northern area of Bern is
consistent with an expectation that political boundaries are used for
inter-religious rather than inter-lingual conflict, for which purpose
they may not be as well adapted.

Specific events, listed by location:

Glovelier - March 24, 1961, arson against a military arsenal. [H:1];
July 16, 1972, explosion of a military arsenal. [H:2]
(\url{http://www.bijube.ch/page-7207.html})

Les Auges - October 21, 1962, arson against a military
barracks. [H:3] (\url{http://www.bijube.ch/page-6210.html})

Bourrignon - March 26, 1963, arson against a military barracks. [H:3]
(\url{http://www.bijube.ch/page-6303.html})

Genevez - April 28, 1963, arson against a farm. [H:3] (\url{http://www.bijube.ch/page-6304.html})

Montfaucon - July 18, 1963, arson against a farm. [H:3] (\url{http://www.bijube.ch/page-6307.html})

Mont-Soleil - October 5, 1963, a house bombing against a leader of an
anti-separatist group.  [H:3,4]
(\url{http://www.bijube.ch/page-6310.html})

Malleray - December 23, 1963, a bombing of a property of an
anti-separatist group leader.  [H:3]; October 20, 1987, arson against
a shooting range. [H:5] (\url{http://www.bijube.ch/page-6312.html},
\url{http://www.bijube.ch/page-8509.html})

Studen - February 27, 1964, bombing of a railway line. [H:3,S8:6]
(\url{http://www.bijube.ch/page-6402.html})

Delemont - March 12, 1964, bombing of a branch of the Cantonal Bank of
Berne. [H:3] ; March 4, 1966, government administration building
attacked. [H:1] (\url{http://www.bijube.ch/page-6403.html})

Saignel\'egier - November 20, 1965, arson against a hotel. [H:1]; On
October 1, 1987, explosion of a munitions depot. [H:7]
(\url{http://www.bijube.ch/page-6511.html},
\url{http://www.bijube.ch/page-8710.html})

Mont-Crosin - May 29, 1966, arson against a hotel. [H:1]
(\url{http://www.bijube.ch/page-6605.html})

Cort\'ebert - March 16, 1980, violent fighting between separatists and
anti-separatists with stones, firecrackers, and flare
guns. Demonstrators on both sides were injured. [H:8]
(\url{http://www.bijube.ch/page-8003.html})

Moutier - September 4, 1985 bombing of the district court. [H:9]
(\url{http://www.bijube.ch/page-8509.html})

Reussilles - September 11 and 23, 1993, arson against a munitions
depot. [H:7,H:5]

Perrefitte - October 21, 1987, bombing of a shooting range. [H:5]
(\url{http://www.bijube.ch/page-8710.html})

B\"uren - April 5, 1989, arson against a historic wooden
bridge. [H:10] (\url{http://www.bijube.ch/page-8509.html})

Montbautier - May 24, 1992, arson against a German-language school,
previously vandalized.  [H:11]

Courtelary - January 7, 1993, bombing of a house of an
anti-separatist. [H:12,S8:13]
(\url{http://www.bijube.ch/page-9301.html}

Berne - January 7, 1993, premature explosion of a bomb in a car
killing one person. [H:13]
(\url{http://www.bijube.ch/page-9301.html})\\

\noindent \textbf{References:}

[H:1] ``Les incendiaires ont avou\'e,'' Journal de Genève, July 6,
1966.

[H:2] ``Le FLJ fait sauter un d\'ep\^ot de munitions de l'arm\'ee,''
Journal de Genève, July 17, 1972.

[H:3] ``De l'incendie \`a l'explosion en passant par les menaces,''
Gazette de Lausanne, October 13, 1965.

[H:4] ``Plastic contre la villa d'un dirigeant UPJ,'' Gazette de
Lausanne, October 7, 1963.

[H:5] ``La s\'erie des attentats s'allonge,'' Journal de Genève,
October 22, 1987.

[H:6] ``Le Sabotage du FLJ serait confirm\'e,'' Gazette de Lausanne,
March 4, 1964.

[H:7] ``D\'ep\^ot de munitions d\'etruit \`a Saignel\'egier,''
Journal de Genève, October 2, 1987.

[H:8] ``Cort\'ebert: violents incidents entre s\'eparatists et
pro-bernois,'' Gazette de Lausanne, March 17, 1980.

[H:9] ``Enormes d\'eg\^ats \`a Moutier,'' Gazette de Lausanne,
September 4, 1985.

[H:10] ``Le pont historique de B\"uren incendi\'e,'' Gazette de
Lausanne, April 6, 1989.

[H:11] ``Les flammes ravagent une \'ecole du Jura bernois,''Le
Nouveau Quotidien, May 26, 1992.

[H:12] ``Deux explosions, dont une mortelle dans le canton de
Berne,'' Gazette de Lausanne, January 8, 1993.

[H:13] ``Retour tragique des poseurs de bombes jurassiens,'' Le
Nouveau Quotidien, January 8, 1993.
\newpage
\section{Yugoslavia}
\label{yugoslavia}
\begin{figure}[bm]
\includegraphics[width=13cm]{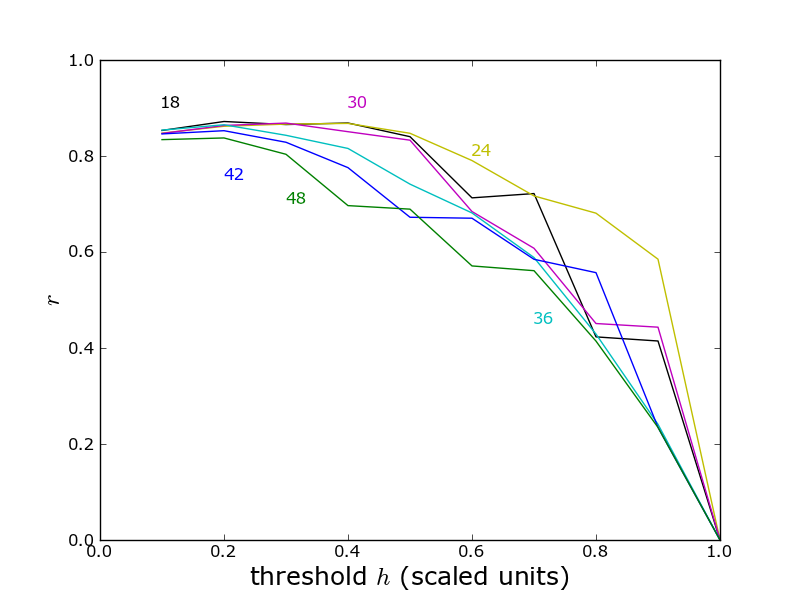}
\caption{\label{figS17} Correlation of proximity maps of predicted and reported
violence in Yugoslavia without topographical or political boundaries,
as a function of threshold for violence divided by the maximum
propensity for violence. Each curve is labelled by the characteristic
length (km).  (Compare with Figure S4.3 in Ref. [14].)}
\end{figure}

\begin{figure}
\includegraphics[width=13cm]{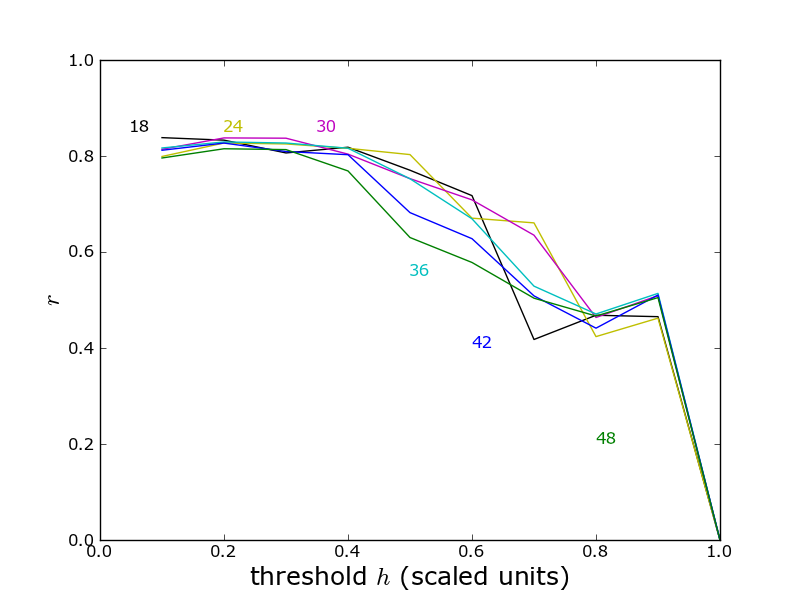}
\caption{\label{figS18} As in Fig. \ref{figS17} but including the
  effects of administrative boundaries.}
\end{figure}

\begin{figure}
\includegraphics[width=13cm]{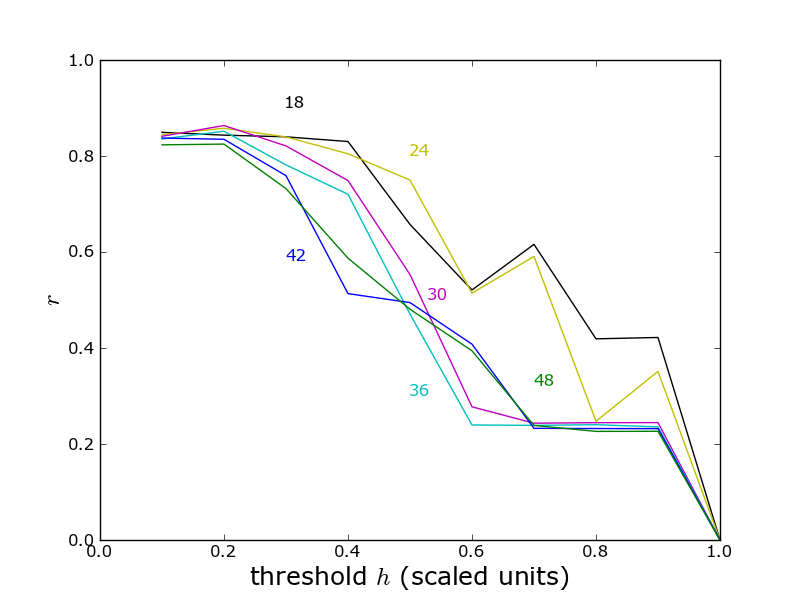}
\caption{\label{figS19} As in Fig. \ref{figS17} but including
  topographical boundaries.}
\end{figure}

Figures showing the correlation of predicted and reported violence for
the former Yugoslavia without administrative or topographical
boundaries (Fig. \ref{figS17}) with administrative boundaries
(Fig. \ref{figS18}) and with topographical boundaries
(Fig. \ref{figS19}).

\begin{figure}
\includegraphics[width=13cm]{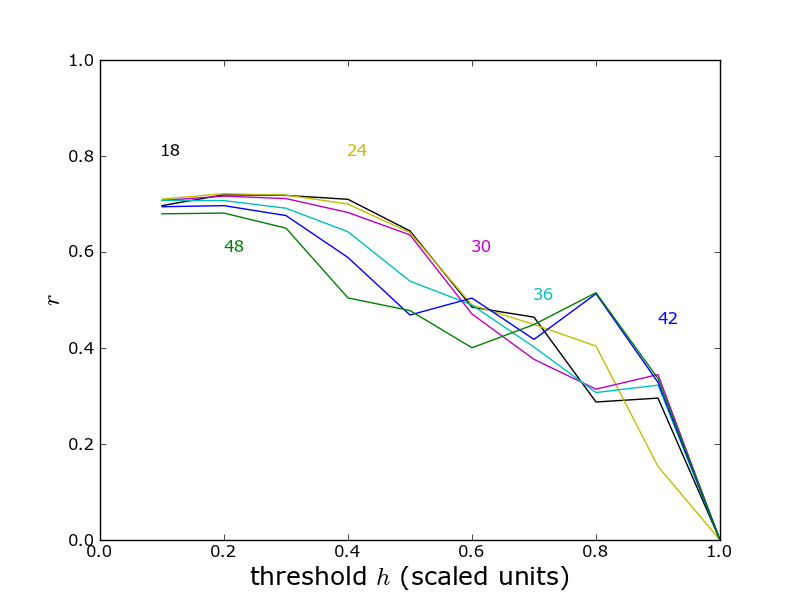}
\caption{\label{figS20} As in Fig. \ref{figS17}, but including
  Slovenia and Macedonia.}
\end{figure}

\begin{figure}
\includegraphics[width=13cm]{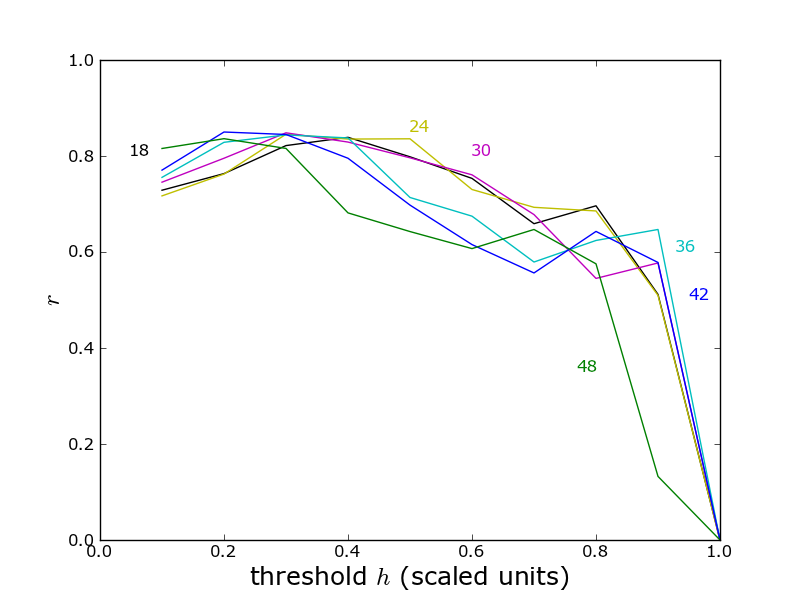}
\caption{\label{figS21} As in Fig. \ref{figS20} but including the
  effect of political boundaries.}
\end{figure}

We also provide a similar analysis of the former Yugoslavia including
Macedonia and Slovenia, without (Fig. \ref{figS20}) and with
(Fig. \ref{figS21}) political boundaries. Without political boundaries
the agreement of predicted and reported violence is dramatically
reduced.

\clearpage

\section{Expanded bibliography on ethnic conflict} 
\label{expanded bibliography}
In recent years there have been increasing efforts to understand the
causes and enabling conditions for civil war and ethnic conflict. The
attached bibliography [J:1--118] expands on citations included in the
main text and in supplementary materials of Ref. [14]. These efforts
include examinations of geography and other structures within
countries [J:18--50] as well as the effects of transnational
geography [J:51--67]. Extensive analysis explores the role of
political structures, particularly federalism, in enabling or
preventing civil and ethnic conflict [J:68--89]. Research has begun
to include quantitative studies and modeling to understand human
behavior and conflict [J:90--94]. A body of research examines
Switzerland regarding the presence or absence of tensions and possible
causes [J:95--118].\\

\noindent \textbf{References:}

[J:1] P. Collier, A. Hoeffler, ``Greed and
grievance in civil war,'' Policy Research Working Paper 2355, The
World Bank Development Research Group (2000).  

[J:2] P. Collier,
A. Hoeffler, D. Rohner, Beyond greed and grievance: feasibility and
civil war.  Oxford Econ. Papers 61, 1, 1--27 (2009).  

[J:3] P. Collier, D. Rohner, Democracy, Development, and
Conflict. J. European Econ. Assoc.  6, 2--3, 531--540 (2008).  

[J:4] K. Eck, L. Hultman, One-sided violence against civilians in
war: Insights from new fatality data. J. Peace Res. 44, 2, 233--246
(2007).

[J:5] J. Esteban, D. Ray, Polarization, fractionalization and
conflict. J. Peace Res. 45, 2, 163--182 (2008).  

[J:6] J. Esteban, G. Schneider, Polarization and conflict:
Theoretical and empirical issues. J.  Peace Res. 45, 2, 131--141
(2008).

[J:7] T. R. Gurr, Ethnic warfare on the wane. Foreign Affairs 79, 3,
52--64 (2000).

[J:8] J. G. Kellas, The Politics of Nationalism and Ethnicity
(St. Martin's Press, New York, NY, ed. 2, 1998).

[J:9] I. Lustick, Stability in deeply divided societies:
Consociationalism versus control.  World Politics 31, 3, 325--344
(1979).

[J:10] J. McGarry, B. O'Leary, Eds., The Politics of Ethnic Conflict
Regulation: Case Studies of Protracted Ethnic Conflicts (Routledge,
New York, NY, 1993).

[J:11] A. Rabushka, K. A. Shepsle, Politics in Plural
Societies: A Theory of Democratic Instability (Merrill, Columbus, OH,
1972).  

[J:12] N. Rost, G. Schneider, J. Kleibl, A global risk
assessment model for civil wars. Social Sci. Res. 38, 4, 921--933
(2009).  

[J:13] S. M. Saideman, D. J. Lanoue, M. Campenni, S. Stanton,
Democratization, political institutions, and ethnic conflict: A pooled
time-series analysis, 1985--1998. Comparative Political Studies 35, 1,
103--129 (2002).

[J:14] N. Sambanis, What
is civil war? Conceptual and empirical complexities of an operational
definition. J. Conflict Res. 48, 6, 814--858 (2004).  

[J:15]
J. Snyder, From Voting to Violence: Democratization and Nationalist
Conflict (Norton, New York, NY, 2000).  

[J:16] E. Spolaore, Civil
conflict and secessions. Economics of Governance 9, 1, 45--63 (2008).

[J:17] A. Wimmer, Who owns the state? Understanding ethnic conflict
in post-colonial societies. Nations and Nationalism 3, 4, 631--665
(1997).  

[J:18] A. Alesina, R. Baqir, C. Hoxby, Political
jurisdictions in heterogeneous communities.  J. Political Economy 112,
2 (2004).  

[J:19] R. Bhavnani, D. Miodownik, Ethnic polarization,
ethnic salience, and civil war. J.  Conflict Res. 53, 1, 30--49 (2009).

[J:20] R. Bhavnani, D. Miodownik, J. Nart, REsCape: An Agent-Based
Framework for Modeling Resources, Ethnicity, and
Conflict. J. Artif. Societies and Social Sim. 11, 27, (2008).

[J:21] R. J. Blimes, The indirect effect of ethnic heterogeneity on
the likelihood of civil war onset. J. Conflict Res. 50, 4, 536--547
(2006).  

[J:22] A. Braithwaite, MIDLOC: Introducing the Militarized
Interstate Dispute Location dataset. J. Peace Res. 47, 1, 91--98
(2010).  

[J:23] H. Buhaug, L. E. Cederman, J. K. Rød, ``Modeling Ethnic
Conflict in Center-Periphery Dyads,'' Paper prepared for presentation
at the workshop on ``Polarization and Conflict,'' Nicosia, Cyprus,
April 26--29, 2006.

[J:24] H. Buhaug, S. Gates, The Geography of Civil War. J. Peace
Res. 39, 4, 417--433 (2002).

[J:25] H. Buhaug, J. K. Rød, Local determinants of African civil
wars, 1970--2001. Polit.  Geog. 25, 3, 315--335 (2006).

[J:26] L. E. Cederman, ``Articulating the geo-cultural logic of
nationalist insurgency,'' in Order, Conflict, and Violence,
S. N. Kalyvas, I. Shapiro, T. Masoud, Eds. (Cambridge University
Press, Cambridge, 2008) pp. 242--270.

[J:27] L. E. Cederman, L. Girardin, Beyond fractionalization:
Mapping ethnicity onto nationalist
insurgencies. Am. Pol. Sci. Rev. 101, 1, 173--185 (2007).

[J:28] L. E. Cederman, L. Girardin, ``Toward realistic computational
models of civil wars,'' Paper prepared for presentation at the Annual
Meeting of the American Political Science Association, held in
Chicago, August 30--September 2, 2007.

[J:29] L. E. Cederman, L. Girardin, K. S. Gleditsch,
Ethnonationalist triads: Assessing the influence of kin groups on
civil wars. World Politics 61, 3, 403--437 (2009).

[J:30] L. E. Cederman, K. S. Gleditsch, Introduction to Special
Issue on ``Disaggregating Civil War.'' J. Conflict Res. 53, 4, 487--495
(2009).

[J:31] L. E. Cederman, J. K. Rød, N. B. Weidmann, ``Geo-referencing
of ethnic groups: Creating a new dataset,'' Prepared and presented at
the GROW Workshop, Peace Research Institute Oslo (PRIO), February
10--11, 2006.

[J:32] L. E. Cederman, A. Wimmer, B. Min, Why do ethnic groups
rebel? New data and analysis. World Politics 62, 1, 87--119 (2010).

[J:33] S. Chojnacki, N. Metternich, ``Event Data Project on Conflict
and Security (EDACS) in Areas of Limited Statehood,'' Coding Manual
Version 1.3. (published 2007, download: \url{http://
  www.polsoz.fu-berlin.de/polwiss/forschung/international/frieden/publikationen/
  C4_Coding_Manual_1_3.pdf}, November 9, 2010).

[J:34] C. Cioffi-Revilla, M. Rouleau, MASON RebeLand: An Agent-Based
Model of Politics, Environment, and Insurgency. International
Stud. Rev. 12, 1, 31--52 (2010).

[J:35] D. Cunningham, K. S. Gleditsch, I. Salehyan, It takes two: A
dyadic analysis of civil war duration and outcome. J. Conflict
Res. 53, 4, 570--597 (2009).

[J:36] H. Dorussen, ``Introducing PKOLED: A peacekeeping operations
location and event dataset,'' Paper presented at the conference
Disaggregating the Study of Civil War and Transnational Violence,
University of Essex, November 24--25, 2007.

[J:37] M. Findley, P. Rudloff, ``Combatant fragmentation and the
dynamics of civil wars,'' American Political Science Association 2009
Toronto Annual Meeting Paper (published 2009, download:
\url{http://ssrn.com/abstract=1450036}, November 9, 2010).

[J:38] M. Findley, J. K. Young, S. M. Shellman, ``Modeling Dynamic
Violence: Integrating Events, Data Analysis and Agent-Based
Modeling,'' American Political Science Association 2010 Annual Meeting
paper (published 2010, download:
\url{http://ssrn.com/abstract=1642577}, November 9, 2010).

[J:39] A. Geller, S. Moss, Growing qawm: An evidence-driven
declarative model of Afghan power structures. Adv. Complex Syst. 11,
2, 321--335 (2008).

[J:40] A. S. Mahmud, The creation of multi-ethnic nations with or
without a core region.  Public Choice (2010, download:
\url{http://www.springerlink.com/content/vg13k3264q21362r/}, January
18, 2011).

[J:41] H. Meadwell, Spatial models of secession-proofness and
equilibrium size. Quality \& Quantity (2010, download:
\url{http://www.springerlink.com/content/0m1kw18015265504/}, January
18, 2011).

[J:42] C. Raleigh, H. Hegre, Population size, concentration, and
civil war. A geographically disaggregated analysis. Polit. Geog. 28,
4, 224--238 (2009).

[J:43] J. K. Rød, H. Buhaug, ``Civil Wars: Prospects and Problems
with the Use of Local Indicators,'' Proceedings, ScanGIS 2007, 212--225
(2007).

[J:44] S. M. Shellman, Coding disaggregated intrastate conflict:
Machine processing the behavior of substate actors over time and
space. Polit. Anal. 16, 4, 464--477 (2008).

[J:45] J. Tir, Dividing countries to promote peace: Prospects for
long-term success of partitions. J. Peace Res. 42, 5, 545--562 (2005).

[J:46] N. B. Weidmann, ``From sparks to prairie fires: spatial
mechanisms of group mobilization,'' Paper prepared for presentation at
the Annual Meeting of the American Political Science Association in
Chicago, August 30--September 2, 2007 and at the Sixth PanEuropean
Conference on International Relations in Torino, September 12--15,
2007.

[J:47] N. B. Weidmann, ``Geographic group fragmentation and civil
peace,'' Paper presented at the conference Disaggregating the Study of
Civil War and Transnational Violence, University of Essex, November
24--25, 2007.

[J:48] N. B. Weidmann, Geography as motivation and opportunity:
Group concentration and ethnic conflict. J. Conflict Res. 53, 4,
526--543 (2009).

[J:49] N. B. Weidmann, D. Kuse, WarViews: Visualizing and animating
geographic data on civil war. International Stud. Perspectives, 10, 1,
36--48 (2009).

[J:50] W. Yu, D. Helbing, ``Game theoretical interactions of moving
agents,'' in Simulating Complex Systems by Cellular Automata,
vol. 0/2010 of Understanding Complex Systems A. G.  Hoekstra, et
al. Eds. (Springer-Verlag, Berlin Heidelberg, 2010) ch. 10, 219--239.

[J:51] A. T. Bohlken, E. J. Sergenti, Economic growth and ethnic
violence: An empirical investigation of Hindu--Muslim riots in
India. J. Peace Res. 47, 5, 589--600 (2010).

[J:52] A. Braithwaite, Resisting infection: How state capacity
conditions conflict contagion. J.  Peace Res. 47, 3, 311--319 (2010).

[J:53] H. Buhaug, S. Gates, P. Lujala, Geography, rebel capability,
and the duration of civil conflict. J. Conflict Res. 53, 4, 544--569
(2009).

[J:54] H. Buhaug, K. S. Gleditsch, Contagion or confusion? Why
conflicts cluster in space.  International Stud. Quarterly 52, 2,
215--233 (2008).

[J:55] N. F. Campos, V. S. Kuzeyev, On the dynamics of ethnic
fractionalization. Am. J. of Pol. Sci. 51, 3, 620--639 (2007).

[J:56] J. D. Fearon, K. Kasara, D. D. Laitin, Ethnic minority rule
and civil war onset. Am. Pol.  Sci. Rev. 101, 1, 187--193 (2007).

[J:57] E. Forsberg, Polarization and ethnic conflict in a widened
strategic setting. J. Peace Res. 45, 2, 283--300 (2008).

[J:58] D. Gerritsen, Unknown is unloved? Diversity and
inter-population trust in Europe.  European Union Politics 11, 2,
267--287 (2010).

[J:59] K. S. Gleditsch, Transnational dimensions of civil
war. J. Peace Res. 44, 3, 293--309 (2007).

[J:60] H. Hegre, M. Nome, ``Democracy, development, and armed
conflict,'' APSA 2010 Annual Meeting Paper (2010).

[J:61] G. Østby, Polarization, horizontal inequalities and violent
civil conflict. J. Peace Res.  45, 2, 143--162 (2008).

[J:62] G. Østby, R. Nordås, J. K. Rød, ``Regional inequalities and
civil conflict in 21 subSaharan African countries, 1986--2004,''
Prepared for presentation at the workshop Polarization and Conflict,
Nicosia, Cyprus, 26--29 April 2006.

[J:63] M. Ross, Michael, A closer look at oil, diamonds, and civil
war. Annu. Rev. Poli. Sci. 9, 265--300 (2006).

[J:64] H. Strand, ``Onset of armed conflict: A new list for the
period 1946--2004, with applications,'' in Reassessing the Civil
Democratic Peace, Ph.D. thesis, Department of Political Science,
University of Oslo \& Centre for the Study of Civil War, PRIO (2006).

[J:65] N. B. Weidmann, J. K. Rød, L. E. Cederman, Representing
ethnic groups in space: A new dataset. J. Peace Res. 47, 4, 491--499
(2010).

[J:66] A. Wimmer, L. E. Cederman, B. Min, ``Ethnic politics and
violent conflicts, 1946--2005: A configurational approach,'' Paper
presented at the conference Disaggregating the Study of Civil War and
Transnational Violence, University of Essex, November 24--25, 2007.

[J:67] A. Wimmer, L. E. Cederman, B. Min, Ethnic politics and armed
conflict: A configurational analysis of a new global data
set. Am. Soc. Rev. 74, 2, 316--337 (2009).

[J:68] E. Aleman, D. Treisman, ``Fiscal politics in ‘ethnically
mined,' developing, federal states: central strategies and
secessionist violence,'' in Sustainable Peace: Power and Democracy
After Civil Wars, P. G. Roeder, D. Rothchild, Eds. (Cornell University
Press, Ithaca, 2005) 173--216.

[J:69] U. M. Amoretti, N. Bermeo, Eds. Federalism and Territorial
Cleavages (Johns Hopkins University Press, Baltimore, MD, 2004).

[J:70] K. M. Bakke, ``Peace-promoting versus peace-preserving
federalism?''  Paper presented at the annual meeting of the
International Studies Association, San Diego, California, March 22,
2006.

[J:71] K. M. Bakke, E. Wibbels, Diversity, disparity, and civil
conflict in federal states. World Politics 59, 1, 1--50 (2006).

[J:72] D. Brancati, Can federalism stabilize Iraq? The Washington
Quarterly 27, 2, 7--21 (2004).

[J:73] D. Brancati, Decentralization: Fueling the fire or dampening
the flames of ethnic conflict and secessionism?  International
Organization 60, 3, 651--685 (2006).

[J:74] T. Christin, S. Hug, ``Federalism, the geographic location of
groups, and conflict,'' Center for Comparative and International
Studies, ETH Zurich and University of Zurich (2006).

[J:75] F. S. Cohen, Proportional versus majoritarian ethnic conflict
management in democracies. Comp. Political Studies 30, 5, 607--630
(1997).

[J:76] R. D. Congleton, ``A political efficiency case for federalism
in multinational states: Controlling ethnic rent-seeking,'' in
Competition and Structure: The Political Economy of Collective
Decisions: Essays in Honor of Albert Breton, G. Galeotti, P. Slamon,
R. Wintrobe, Eds. (Cambridge University Press, New York, NY, 2000),
pp. 284--308.

[J:77] H. E. Hale, Divided we stand: Institutional sources of
ethnofederal state survival and collapse. World Politics 56, 2,
165--193 (2004).

[J:78] C. A. Hartzell, Explaining the stability of negotiated
settlements to intrastate wars.  J. Conflict Res. 43, 1, 3--22 (1999).

[J:79] S. Hug, ``Conflict and federal arrangements,'' Paper
presented at the annual meeting of the International Studies
Association, San Diego, California, March 22, 2006.

[J:80] C. Johnson, ``Federalism and secession under conditions of
civil war,'' Paper presented at the annual meeting of the
International Studies Association, San Diego, California, March 22,
2006.

[J:81] D. A. Lake, D. Rothchild, ``Territorial decentralization and
civil war settlements,'' in Sustainable Peace: Power and Democracy
After Civil Wars, P. G. Roeder, D. Rothchild, Eds.  (Cornell
University Press, Ithaca, London, 2005) pp. 109--132.

[J:82] J. McGarry, B. O'Leary, ``Federation as a method of ethnic
conflict regulation,'' in From Power Sharing to Democracy:
Post-Conflict Institutions in Ethnically Divided Societies, S.  Noel,
Ed. (McGill-Queen's University Press, Montreal, 2005) ch. 13,
pp. 263--296.

[J:83] E. A. Nordlinger, Conflict Regulation in Divided Societies
(Center for International Affairs, Harvard University, Cambridge, MA,
1972).

[J:84] A. Rabushka, K. A. Shepsle, Politics in Plural Societies: A
Theory of Democratic Instability (Merrill, Columbus, OH, 1972).

[J:85] A. Schou, M. Haug, ``Decentralisation in conflict and
post-conflict situations,'' NIBR Working Paper: 2005:139.

[J:86] E. Spolaore, ``Federalism, regional redistribution, and
country stability,'' Prepared for the 5th Symposium on Fiscal
Federalism ``Regional fiscal flows, balance-sheet federalism, and the
stability of federations,'' IEB-IEA, Barcelona, June 19--20, 2008.

[J:87] A. Stepan, Federalism and democracy: Beyond the
U.S. model. J. Democracy 10, 4, 19--34 (1999).

[J:88] R. L. Watts, Federalism, federal political systems, and
federations. Annu. Rev. Poli. Sci.  1, 117--137 (1998).

[J:89] E. Wibbels, Madison in Baghdad? Decentralization and
federalism in comparative politics. Annu. Rev. Poli. Sci. 9, 165--188
(2006).

[J:90] J. C. Bohorquez, S. Gourley, A. R. Dixon, M. Spagat,
N. F. Johnson, Common ecology quantifies human insurgency. Nature 462,
911--914 (2009).

[J:91] H. Hegre, N. Sambanis, Sensitivity analysis of empirical
results on civil war onset. J.  Conflict Res. 50, 4, 508--535 (2006).

[J:92] L. Luo, N. Chakraborty, K. Sycara, Modeling ethno-religious
conflicts as prisoner's dilemma game in graphs. International
Conference on Computational Science and Engineering 4, 442--449 (2009).

[J:93] I. S. Lustick, D. Miodownik, Abstractions, ensembles, and
virtualizations: simplicity and complexity in agent-based
modeling. Comparative Politics 41, 2, 223--244 (2009).

[J:94] D. R. White, ``Dynamics of human behavior,'' in
          Encyclopedia of Complexity and Systems Science,
          R. A. Meyers, Ed.-in-chief (Springer-Verlag, Berlin,
          Heidelberg, 2009).  

[J:95] C. H. Church, Switzerland: a paradigm in
evolution. Parliam. Aff. 53, 1, 96--113 (2000).

[J:96] J. Dunn, ``Consociational democracy'' and language conflict:
A comparison of the Belgian and Swiss experiences. Comparative
Political Studies 5, 1 (1972).

[J:97] U. D\"urm\"uller, Swiss multilingualism and intranational
communication. Sociolinguistica 5, 111--159 (1991).

[J:98] M. J. Esman, Ed. Ethnic Conflict in the Western World
(Cornell University Press, Ithaca, NY, 1977).

[J:99] H. E. Glass, Ethnic diversity, elite accommodation and
federalism in Switzerland.  Publius 7, 4, Federalism and Ethnicity
31--48 (1977).

[J:100] R. C. Head, Catholics and Protestants in Graub\"unden:
Confessional discipline and confessional identities without an early
modern state?  German History 17, 3, 321--345 (1999).

[J:101] D. L. Horowitz, Ethnic Groups in Conflict (University of
California Press, Berkeley, Los Angeles, CA, 1985).

[J:102] R. S. Katz, Dimensions of partisan conflict in Swiss
cantons. Comp. Political Stud. 16, 4, 505--527 (1984).

[J:103] W. R. Keech, Linguistic diversity and political conflict:
Some observations based on four Swiss cantons. Comp. Politics 4, 3,
387--404 (1972).

[J:104] A. Lijphart, Democracy in Plural Societies: A Comparative
Exploration (Yale University Press, New Haven and London, 1977).

[J:105] W. Linder, Swiss Democracy: Possible Solutions to Conflict
in Multicultural Societies (Macmillan Press, New York, ed. 2, 1998).

[J:106] K. D. McRae, Conflict and Compromise in Multilingual
Societies: Switzerland (Wilfrid Laurier University Press, Waterloo,
Ontario, Canada, 1983).

[J:107] L. Pap, The language situation in Switzerland: An updated
survey. Lingua 80, 2--3, 109--148 (1990).

[J:108] W. E. Rappard, Federalism in Switzerland. Parliam. Aff. IV,
2, 236--244 (1950).

[J:109] L. Ruttan, M. Franzen, R. Bettinger, P. J. Richerson,
``Ethnic Interactions: Analysis of a Sample of Boundaries,'' (2006,
download:
\url{http://www.des.ucdavis.edu/faculty/Richerson/Ethnicity\%208-11-06\%20for\%20pdf.pdf},
November 11, 2010).

[J:110] C. L. Schmid, Conflict and Consensus in Switzerland
(University of California Press, Berkeley and London, 1981).

[J:111] J. Steinberg, Why Switzerland? (Cambridge University Press,
Cambridge, UK, ed. 2, 1996).

[J:112] P. Stevenson, Political Culture and Intergroup Relations in
Plurilingual Switzerland. J.  Multilingual and Multicultural
Development 11, 3, 227--255 (1990).

[J:113] N. Stojanovic, ``The idea of a Swiss nation: A critique of
Will Kymlicka's account of multination states,'' M.A. Thesis, McGill
University Department of Political Science (2000, download:
\url{http://www.nenadstojanovic.ch/testi\%20scientifici/M.A.\%20Thesis\%20-\%20final\%20version.pdf},
November 11, 2010).

[J:114] D. Welsh, Domestic politics and ethnic conflict. Survival
35, 1, 63--80 (1993).

[J:115] A. S. Wilner, The Swiss-ification of ethnic conflict:
Historical lessons in nationbuilding from the Swiss example. Federal
Governance 5/6, 2010).

[J:116] F. Rash, The German-Romance language borders in
Switzerland. J. Multilingual and Multicultural Development 23, 1/2,
112--136 (2002).

[J:117] P. Stevenson, Political culture and intergroup relations in
plurilingual Switzerland. J.  Multilingual and Multicultural
Development 11, 3, 227--255 (1990).

[J:118] O. Zimmer, A Contested Nation: History, Memory and
Nationalism in Switzerland, 1761--1891 (Cambridge University Press,
Cambridge and New York, 2007).

\end{document}